\documentclass[reprint,onecolumn,secnumarabic,amssymb, nobibnotes, aps, prfluids]{revtex4-2}

\usepackage{psfrag}

\usepackage{xcolor}
\usepackage{bbm}

\usepackage{lmodern}
\usepackage[T1]{fontenc}
\usepackage[utf8]{inputenc}
\usepackage{indentfirst}
\usepackage{color}
\usepackage{microtype}
\usepackage{lipsum}	
\usepackage{parskip}
\usepackage{setspace}

\usepackage{graphicx}
\usepackage{subcaption}
\usepackage{epstopdf}
\usepackage{epsfig}
\usepackage{rotating}
\usepackage{morefloats}

\usepackage{amsmath}
\usepackage{amssymb}
\usepackage{bm}
\usepackage{wasysym}
\usepackage{icomma}
\usepackage{multirow}
\usepackage{placeins}
\usepackage{mathrsfs}
\usepackage{mathtools}
\usepackage{subeqnarray}

\usepackage[english]{babel}
\usepackage{bibentry}
\usepackage{hyperref}
\usepackage{cleveref}
\usepackage{natbib}

\usepackage{etoolbox}
\makeatletter
\patchcmd\affils@present@group
 {\frontmatter@affiliationfont}
 {\frontmatter@affiliationfont\begingroup\lccode`~=`,\lowercase{\endgroup\let~}\active@comma}
 {}{}
\makeatother

\usepackage{makecell}

\begin{document}


\title{Coherent pressure structures in turbulent channel flow}


\author{Filipe R. do Amaral}
\thanks{filipe.ramos.do.amaral@univ-poitiers.fr}
\affiliation{D\'{e}partement~Fluides,~Thermique,~Combustion,~Institut~Pprime-CNRS-Universit\'{e}~de~Poitiers-ENSMA,~86000~Poitiers,~France}
\author{Andr\'{e} V. G. Cavalieri}
\thanks{andre@ita.br}
\affiliation{Divis\~{a}o~de~Engenharia~Aeroespacial,~Instituto~Tecnol\'{o}gico~de~Aeron\'{a}utica,~12228-900~S\~{a}o~Jos\'{e}~dos~Campos,~SP,~Brazil}
\date{\today}
\maketitle


Most of the studies on pressure fluctuations in wall-bounded turbulent flows aim at obtaining statistics as power spectra and scaling laws, especially at the walls.
In the present study we study energetic coherent pressure structures of turbulent channel flows, aiming at a characterization of dominant coherent structures throughout the channel.
Coherent structures are detected using spectral proper orthogonal decomposition (SPOD) and modeled using resolvent analysis, similar to related works dealing with velocity fluctuations, but this time using pressure fluctuations as the output of interest.
The resolvent operator was considered with and without the Cess eddy viscosity model.
Direct numerical simulations (DNSs) of incompressible turbulent channel flows at friction Reynolds numbers of approximately 180 and 550 were employed as databases in this study. 
Three representative dominant structures emerged from a preliminary spectral analysis: near-wall, large-scale and spanwise-coherent structures.
For frequency-wavenumber combinations corresponding to these three representative structures, SPOD results show a strong dominance of the leading mode, highlighting low-rank behavior of pressure fluctuations.
The leading resolvent mode closely agrees with the first SPOD mode, providing support to studies that showed better performance of resolvent-based estimators when predicting pressure fluctuations compared to velocity fluctuations (Amaral \emph{et al., J. Fluid Mech.}, vol. 927, A17, 2021).
The dominant mechanisms of the analyzed modes are seen to be the generation of quasi-streamwise vortices with pressure fluctuations appearing close to vortex centers.
A study on the individual contributions of the nonlinear terms (treated as forcing in resolvent analysis) to the pressure output reveals that each forcing component plays a constructive role to the input-output formulation, which also helps understanding the weaker role of forcing ``color'' in driving pressure fluctuations.

\section{Introduction}
\label{sec:introduction}

The understanding of pressure fluctuations in wall-bounded turbulent flows is of fundamental importance for many practical applications such as sound radiation and structural vibration.
With the world community appeal for ``greener by design'' vehicles and technologies, lower sound emission is one of the goals that have been pursued towards diminishing environmental pollution, as noise is also a concern regarding human and wild life health issues.
On the other hand, fluid-structure interactions are of concern for structural integrity.
Moreover, pressure fluctuations on the surface of a sharp body, such as the trailing-edge of a wing, scatters sound that is propagated to the far-field and may also induce structural vibrations.
Besides aircraft wings, this is especially important for wind-turbines, fans and propellers \citep{lee2021turbulent}.

Although the last few decades have experienced great advances in both computations and experiments of wall turbulence, the study of pressure fluctuations in wall-bounded turbulent flows has received less attention than the analysis of velocity fluctuations.
On the experimental side, most of the studies were focused on the analysis of wall pressure measurements \citep{willmarth1975pressure, schewe1983structure, farabee1991spectral, bull1996wall}, with the microphones usually mounted in arrays flush to the test model wall \citep{arguillat2010measured}.
Such arrays have limited aperture and a finite number of sensors, which, among other issues, limits the measurement resolution.
Test facilities also suffer from background noise that can contaminate the measurements \citep{gravante1998characterization}, especially at low Reynolds numbers \citep{tsuji2007pressure}.
The small scales associated with the turbulent flow also impose a challenge to reliable measurement of pressure fluctuations \citep{klewicki2008statistical}.
Hence, an important part of the literature regarding such topic relies on theoretical and numerical studies.

High-fidelity numerical simulations have the advantage of obtaining the entire flow field, not necessarily limited to the channel walls.
For instance, \citet{kim1987turbulence} conducted a direct numerical simulation (DNS) of a channel flow with friction Reynolds number ($Re_{\tau}$) of 180.
Their database was further used by \citet{choi1990space}, who computed the wall pressure spectra.
When comparing their frequency spectra with experimental data at different Reynolds numbers, the authors obtained different scaling laws for the low- and high-frequency ranges.
The lower frequencies better collapsed with outer variables, whereas inner variables better fitted the higher frequencies.
This result agrees with the experimental data by \citet{farabee1991spectral}, who studied wall pressure fluctuations in boundary layers.
\citet{choi1990space} also identified a small region with a -5 power law decay rate for both frequency and streamwise wavenumber spectra, i.e. for $\omega \nu / {u_{\tau}}^2 > 1$, where $\omega$ is the angular frequency, $\nu$ is the fluid kinematic viscosity and $u_{\tau}$ is the friction velocity, and for $\alpha^+ = \alpha \nu / u_{\tau} > 0.1$, where $\alpha$ is the streamwise wavenumber.
Such decay rate was also found in later studies \citep{hu2006wall, anantharamu2020analysis, yang2022wavenumber}.
For even higher frequencies, the spectra decay at higher rates.
Another observation by \citet{choi1990space} regards the effect of the Reynolds number on the wall pressure spectra.
Increasing the Reynolds number leads to increased power at the high-frequency band when outer scaling is considered, whereas the low-frequency band increases the power with Reynolds number for inner scaling.

\citet{kim1989structure} also explored the $Re_{\tau} = 180$ database by \citet{kim1987turbulence}, focusing on the pressure fluctuations.
The author decomposed the pressure fluctuations into rapid (linear) and slow (nonlinear) components, where the distinction between such components relies on the linear and nonlinear interactions present in the Poisson equation for pressure.
The rapid part of pressure fluctuations, which contains the mean velocity gradient, responds immediately to any change imposed to the flow field, whereas the slow part reacts through nonlinear interactions.
The results obtained by \citet{kim1989structure} show that the slow component is dominant over the rapid component, i.e. the nonlinear source terms in the Poisson equation play a major role in the pressure fluctuations behavior.
Another relevant finding by the author was obtained through the modeling of the wall fluctuations by a Green's function, where the pressure was written as a convolution of a Green's function with a source term.
He showed the instantaneous pressure at one of the channel walls is affected by source terms that are present at the opposite wall, highlighting that pressure fluctuations comprise large spatial scales and have a non-local nature.

Regarding the power spectra of wall pressure and shear stress, \citet{hu2006wall} conducted a series of numerical simulations for channel flows up to $Re_{\tau} = 1440$.
The authors compared their results with spectra of channel, pipe and boundary-layer flows at several Reynolds numbers.
The wall pressure spectrum showed a reasonable collapse for $Re_{\tau} \geq 360$, with inner scaling for the higher frequencies and outer scaling for the lower frequencies, as previously observed by \citet{choi1990space}.
The premultiplied frequency spectra exhibited a peak at $\omega^+ \approx 0.3$.
This peak frequency is in agreement with the results obtained by \citet{anantharamu2020analysis}, which conducted DNS of channel flow at $Re_{\tau} = 180$ and 400 aiming at exploring the source of wall pressure fluctuations.
\citet{anantharamu2020analysis} also found that the wall pressure fluctuation peak for the premultiplied wavenumber spectra corresponds to a streamwise wavelength ${\lambda_{x}}^+ \approx 200$, where ${\lambda_{x}}^+ = 2 \pi / \alpha^+$, for fluctuations in the buffer layer for $y^+ \approx 15$, where $y^+$ is the wall-normal coordinate in inner (wall/viscous) units and $\alpha^+$ is the streamwise wavenumber in inner units.
A pressure premultiplied spectrum peak in the range ${\lambda_{x}}^+ \approx 200$ to 300 was also reported by \citet{panton2017correlation}, who studied several DNS databases of channel flow for $Re_{\tau}$ up to 5186, and zero pressure gradient boundary-layer flows for $Re_{\tau}$ up to 1271.
In addition, \citet{panton2017correlation} proposed scaling laws for the pressure statistics at the wall, inner and outer layers.

In recent years there has been substantial effort in modeling coherent structures in wall turbulence \citep{sharma2013coherent, cossu2017self, symon2021energy}.
Efforts towards this direction include the study of linearized models and modal decomposition of databases \citep{moarref2013model, zare2017colour, mckeon2017engine}.
For linearized models, the problem is written in the input-output format in the frequency domain and the nonlinear terms of the Navier-Stokes (N--S) system are treated as an external forcing that generates a response through the resolvent operator, obtained by inversion of the linearized N--S (LNS) operator considering the mean profile as a base flow \citep{mckeon2010critical, hwang2010linear}.
The inputs correspond to the forcing terms (nonlinear terms of N--S system), whereas the outputs denote the flow state (e.g. velocity and pressure components).
In frequency domain, the input and outputs are written as two-point space-time statistics, i.e. cross-spectral densities (CSDs), and can be accurately computed directly from, e.g. snapshots of DNS data.
Resolvent analysis permits the elucidation of the role played by the nonlinear terms of the N--S system.
Application of singular value decomposition (SVD) on the resolvent operator provides the most amplified forcing and response modes.
Resolvent modes also display an hierarchical character, ranked from high to low gain.
It should be emphasized that in order to obtain the resolvent operator, hence the resolvent modes, only the mean flow and the boundary conditions are needed.
As the resolvent analysis employs a linearization over a mean flow, it may reveal the dominant amplification mechanisms.
This is in contrast with the use of the Poisson equation for pressure fluctuations, where no assumption is made about the base flow, and a normal linear operator (the Laplacian) does not display gain separations as significant as the ones obtained with the LNS.
For instance, \citet{luhar2014structure} used resolvent analysis to explore wall pressure fluctuations in turbulent pipe flow obtained through DNS and verified that the rank-1 (optimal) response mode contains most of the relevant features present in the DNS.
They also show the pressure fluctuations that can be identified through the resolvent analysis are related to the rapid (linear) pressure component.
The slow (nonlinear) pressure component, on the other hand, needs the solution of the full nonlinear turbulence field, since it is related to the divergence of the full forcing field.

Response modes obtained from resolvent analysis may be compared to appropriate modal decompositions of data, and a suitable approach is spectral proper orthogonal decomposition (SPOD).
SPOD extracts a set of orthogonal modes from flow realizations in frequency domain that are optimal according to an appropriate inner product \citep{towne2018spectral, schmidt2019efficient}.
The modes are ranked according to their energetic content, i.e. the first mode is the most energetic one and the subsequent modes contain less energy than the previous mode.
The resolvent/response modes are equivalent to the SPOD modes when the flow is forced by stochastic white-noise \citep{towne2018spectral, lesshafft2019resolvent, cavalieri2019wave}.
Moreover, if the resolvent has a large gain separation, i.e. $\sigma_1 \gg \sigma_{2,3,\cdots,n}$, the optimal flow response is expected to be dominant, and is thus similar to the most energetic SPOD mode.
In this case, even a white-noise assumption for the forcing terms in the resolvent framework leads to reasonable agreement between SPOD and resolvent analysis.
Therefore, the resolvent analysis becomes a powerful method to predict dominant coherent structures in turbulent flows.

SPOD and resolvent analysis applied to study coherent structures in wall-bounded flows normally focus on the velocity components.
Early works by \citet{mckeon2010critical} and \citet{hwang2010linear} investigated pipe and channel flows, respectively, for a wide range of Reynolds numbers using resolvent analysis to study the most amplified velocity responses. 
In the channel flow problem studied by \citet{hwang2010linear}, the authors showed the most amplified structures are streamwise streaks which are forced by streamwise vortices.
Streaks are coherent structures, usually identified in the near-wall region of wall-bounded turbulent flows and can be pictured as spanwise aligned layers of low- and high-speed streamwise velocity flow driven by quasi-streamwise vortices that organize the momentum transfer and energy dissipation, redistributing the kinetic energy \citep{farrell2012dynamics, jimenez2013near}.
\citet{abreu2020resolvent} studied DNS channel flow databases at $Re_{\tau} \approx 180$ and 550 and employed SPOD and resolvent analysis to identify and model near-wall coherent structures for several combinations of wavenumbers and frequencies, with only velocity state components considered in the study.
They verified that streaks and streamwise vortices were the most energetic structures in the near-wall region, as expected from the literature of wall-bounded turbulent flows \citep{jimenez2013near}.
The leading SPOD and resolvent modes were in close agreement when the lift up mechanism \citep{ellingsen1975stability, landahl1980note} was active, similarly to what was found in a related turbulent pipe flow study \citep{abreu2020spod}.
Lift up is a linear mechanism that amplifies finite amplitude streamwise fluctuations in shear flows through the action of vertical fluctuations, hence moving low-speed flow from the walls upwards \citep{brandt2014lift, jimenez2018coherent}.

The resolvent operator can be written to include an eddy viscosity model (which substitutes the constant molecular viscosity) to mimic a portion of the nonlinear terms of the N-S system \citep{hwang2010linear, mckeon2017engine}.
In this approach, the nonlinear forcing statistics are still modeled as white noise but the linear operator is modified by the inclusion of the eddy viscosity model, the most common model being the one proposed by \citet{cess1958survey}.
Eddy viscosity approaches were successfully employed, e.g. to compute optimal temporal transient growths supported in turbulent channel flow \citep{delalamo2006linear, pujals2009note}, optimally amplified streaks \citep{butler1993optimal}, to introduce optimal perturbations (streaks) in confined turbulent flows \citep{hwang2010amplification}, among others.
\citet{symon2021energy} and \citet{symon2023eddy} observed the use of a Cess eddy viscosity model was helpful in the resolvent analysis in channel flows, improving the resolvent modes capacity to predict the most energetic structure (streamwise streaks).
However, the use of a singe eddy viscosity profile does not lead to accurate predictions at all scales.

The work by \citet{morra2021colour} employed SPOD and resolvent analysis to investigate turbulent channel flow at $Re_{\tau} \approx 180$ and 550 obtained through DNS.
The authors focused on extracting the nonlinear forcing terms from the database in order to show how such terms affect the resolvent outputs.
In their study, the power spectral density (PSD) of the state (velocity fluctuation) components, evaluated through the input-output formulation for different forcing statistics (input) models, where compared with the true state components, obtained directly from the DNS, for the most energetic near-wall and large-scale structures.
They showed that a rank-2 approximation of the forcing, containing only the two most energetic SPOD modes, is a sufficient approximation of the nonlinear terms, leading to a response with reasonable accuracy when comparing with DNS data.
The performance of the Cess eddy viscosity-modeled forcing on the accuracy of resolvent analysis was also explored and the authors verified the use of the eddy viscosity model boosted the accuracy of the predictions of the output (response) in comparison to the resolvent operator using only the molecular viscosity.

The present study explores the DNS databases of turbulent channel flow at $Re_{\tau} \approx 180$ and 550.
Such DNS databases were validated against established results in the literature.
The aim is to investigate coherent pressure fluctuations in channel flows using SPOD and resolvent analysis.
The coherent structures were obtained through the analysis of the pressure spectra at $y/H = 0$ (corresponding to the channel wall) and $y/H = 0.5$, where $H$ is the channel half-height, targeting the most energetic structures.
Such wall-normal distances were selected as representative locations of near-wall and large-scale structures in wall-bounded flows.
The resolvent operator is studied with and without the inclusion of the Cess eddy viscosity model \citet{cess1958survey}.
We examine the accuracy of input-output analysis to predict pressure fluctuations and the role of nonlinear forcing statistics in constructing the flow response, in order to provide knowledge on the dynamics of pressure structures in wall turbulence.
Low-rank models for the forcing component and their effect on the pressure component estimate is also addressed.

The remainder of this paper is organized as follows.
After this introduction, \S \ref{sec:methodology} addresses the methodology, including the description of the physical problem, the tools employed to analyze the data (resolvent analysis / input-output formulation, two-point statistics and SPOD), and details of the DNS databases.
The energy spectra as a function of the streamwise and spanwise wavenumbers, as well as the frequency, for wall-normal distances corresponding to the near-wall and large-scale structures are reported in \S \ref{sec:energy_spectra}.
Comparisons between SPOD and resolvent modes, with and without the inclusion of the Cess eddy viscosity model on the linear operator are discussed in \S \ref{sec:spod_resolvent_analysis}.
In \S \ref{sec:input_output_analysis} we employ the input-output formulation to elucidated the role of the forcing terms through the study of the effect of each forcing component (input) on the pressure component (output).
Finally, \S \ref{sec:conclusions} summarizes the conclusions.

\section{Methodology}
\label{sec:methodology}

\subsection{Governing equations}
\label{sec:governing_equations}

The flow is governed by the incompressible Navier-Stokes system in the absence of external body forces.
The vector of velocity components is given by $\boldsymbol{u} = \left[u_x~u_y~u_z\right]$, with $u_x$, $u_y$ and $u_z$ as the streamwise, wall-normal and spanwise velocity components, respectively.
The full state vector is written as
\begin{equation}
	\boldsymbol{q} = \left[\boldsymbol{u}~p\right]^{T} \mbox{,}
	\label{eq:state_components}
\end{equation}
\noindent with $\boldsymbol{q} = \boldsymbol{q}\left(x,y,z,t\right)$, where $x$, $y$ and $z$ indicate the streamwise, wall-normal and spanwise directions, respectively, $t$ denotes time, $p$ is the pressure component and $T$ denotes transpose operator.
A Reynolds decomposition was employed over the flow state components, i.e. $\boldsymbol{q} = \boldsymbol{\bar{q}} + \boldsymbol{q^{\prime}}$, were $\boldsymbol{\bar{q}}$ is the mean flow and $\boldsymbol{q^{\prime}}$ is the fluctuation.

In the present study, the mean flow at the streamwise direction is employed to construct the linearized Navier-Stokes system, e.g. $\boldsymbol{\bar{q}} = [\boldsymbol{\bar{U}}(y)~\boldsymbol{z}~\boldsymbol{z}~\boldsymbol{z}]^T$, where $\boldsymbol{z}$ is the zeros vector and $\boldsymbol{\bar{U}}(y)$ is the mean flow velocity profile.
The mean pressure gradient that drives the channel flow has no contribution for the linearised Navier-Stokes system for non-zero frequency and wavenumber, as discussed by \citet{morra2021colour}, and hence is not considered here.
Following \citet{morra2021colour} and dropping the primes from the fluctuation quantities to simplify notations, the Reynolds-decomposed N--S system is given by
\begin{subeqnarray}
	\partial_t \boldsymbol{u} + \left(\boldsymbol{u} \cdot \boldsymbol{\nabla}\right) \boldsymbol{\bar{U}} + \left(\boldsymbol{\bar{U}} \cdot \boldsymbol{\nabla}\right) \boldsymbol{u} + \boldsymbol{\nabla} p - \frac{1}{Re_{\tau}} {\boldsymbol{\nabla}^2 \boldsymbol{u}} &=& \boldsymbol{f}\mbox{,} \\
	\boldsymbol{\nabla} \cdot \boldsymbol{u} &=& 0\mbox{,}
	\label{eq:N-S_system}
\end{subeqnarray}
\noindent where $\partial_t$ denotes partial derivative with respect to time $t$, $\boldsymbol{\nabla}$ is the gradient operator and $\boldsymbol{\nabla}^2$ is the Laplacian operator.
The nonlinear terms of the N--S system, i.e.
\begin{equation}
	\boldsymbol{f} = - \left(\boldsymbol{u} \cdot \boldsymbol{\nabla}\right) \boldsymbol{u} + \overline{\left(\boldsymbol{u} \cdot \boldsymbol{\nabla}\right) \boldsymbol{u}} \mbox{,}
	\label{eq:forcing_components}
\end{equation}
\noindent are interpreted as external forcing \citep{mckeon2010critical} and, in Cartesian coordinates, are given by the vector $\boldsymbol{f} = [f_x~f_y~f_z]^T$, where $f_x$, $f_y$ and $f_z$ are the streamwise, wall-normal and spanwise forcing components, respectively.
The friction Reynolds number is defined as $Re_{\tau} = u_{\tau} H / \nu$.
The Reynolds number may also be given in terms of the bulk velocity $U_b$, i.e. $Re = U_b H / \nu$.

\subsection{Resolvent analysis / input-output formulation}
\label{sec:resolvent_modeling}

Resolvent analysis is a technique in which the governing equations are linearized over a base flow and written in the input-output format.
The inputs refer to the forcing terms and the outputs are the flow responses \citep{jovanovic2005componentwise, mckeon2010critical, hwang2010linear}.
The LNS operator includes the base flow and adequate boundary conditions, e.g. no-slip conditions at the top and bottom walls for channel flows.

For a discrete domain, one has
\begin{subeqnarray}
	\boldsymbol{M} \frac{d \boldsymbol{q}(t)}{d t} &=& \boldsymbol{A} \boldsymbol{q}(t) + \boldsymbol{B} \boldsymbol{f} \mbox{,} \\
	\boldsymbol{y}(t) &=& \boldsymbol{C} \boldsymbol{q}(t)\mbox{,}
	\label{eq:state-space_time}
\end{subeqnarray}
\noindent where $\boldsymbol{M}$ is a diagonal matrix whose entries are set to one and zero for the momentum and continuity equations, respectively, $\boldsymbol{A}$ is the linear operator, assumed as stable, $\boldsymbol{B}$ is the input matrix that restricts the forcing terms to appear only in the momentum equation, and $\boldsymbol{C}$ is a matrix that selects a relevant output $\boldsymbol{y}$.
Throughout this study, $\boldsymbol{C}$ restricts the observations to the components of the state related to pressure fluctuations, unless otherwise specified.
Expression \ref{eq:state-space_time} is exact provided that the forcing terms are computed from the same DNS database employed to obtain the state components.

Equation \ref{eq:state-space_time} can be written in frequency and wavenumber domains as
\begin{equation}
	\boldsymbol{\hat{y}}(\omega) = \left[\boldsymbol{C} \left(- i \omega \boldsymbol{M} - \boldsymbol{A}(\omega)\right)^{-1} \boldsymbol{B}\right] \boldsymbol{\hat{f}}(\omega) \mbox{,}
	\label{eq:state-space_freq}
\end{equation}
\noindent where dependencies on streamwise and spanwise wavenumbers $\alpha$ and $\beta$, respectively, and wall-normal distance $y$, were dropped to simplify notations.
Moreover,
\begin{equation}
	\boldsymbol{\hat{y}}(\omega) = \int \limits_{-\infty}^{\infty} \boldsymbol{y}(t) e^{i \omega t} dt
	\label{eq:fourier}
\end{equation}
is the Fourier transform of $\boldsymbol{y}$, with analogous expression for other variables.
Hats indicate variables in the frequency domain, $\omega$ is the temporal frequency and $i = \sqrt{-1}$.

In expression \ref{eq:state-space_freq}, between brackets one has $\boldsymbol{H} = \boldsymbol{C} \boldsymbol{R} \boldsymbol{B}$, where $\boldsymbol{R} = \boldsymbol{L}^{-1}$ and $\boldsymbol{L} = \left(- i \omega \boldsymbol{M} - \boldsymbol{A}\right)$.
$\boldsymbol{L}$ is the linearized Navier-Stokes operator (LNS), whereas $\boldsymbol{R}$ denotes the resolvent operator and $\boldsymbol{H}$ is the resolvent operator including, through operators $\boldsymbol{C}$ and $\boldsymbol{B}$, the observation and forcing restrictions, respectively.

Following \citet{mckeon2017engine}, the LNS operator is written as 
\begin{equation}
	\boldsymbol{L} =
	\left[\begin{array}{cccc}
		i \alpha \boldsymbol{\bar{U}} - i \omega \boldsymbol{I} - \frac{1}{Re} \boldsymbol{\nabla^2}	& \boldsymbol{\frac{d \bar{U}}{d y}}	& \boldsymbol{Z}	& i \alpha \boldsymbol{I} \\
		\boldsymbol{Z} & i \alpha \boldsymbol{\bar{U}} - i \omega \boldsymbol{I} - \frac{1}{Re} \boldsymbol{\nabla^2}	& \boldsymbol{Z}	& \boldsymbol{\frac{d}{d y}} \\
		\boldsymbol{Z} & \boldsymbol{Z}	& i \alpha \boldsymbol{\bar{U}} - i \omega \boldsymbol{I} - \frac{1}{Re} \boldsymbol{\nabla^2}	& i \beta \boldsymbol{I} \\
		i \alpha \boldsymbol{I}	& \boldsymbol{\frac{d}{d y}}	& i \beta \boldsymbol{I}	& \boldsymbol{Z}
	\end{array}\right] \mbox{,}
	\label{eq:linear_operator}
\end{equation}
\noindent for a triplet $(\alpha,~\beta,~\omega)$.
Here, $\boldsymbol{\bar{U}}$ is the mean turbulent streamwise velocity profile, $\boldsymbol{\nabla^2} = \left(\boldsymbol{\frac{d^2}{d y^2}} - k^2 \boldsymbol{I}\right)$ with $k^2 = \alpha^2 + \beta^2$, $\boldsymbol{\frac{d}{d y}}$ and $\boldsymbol{\frac{d^2}{d y^2}}$ are, respectively, the first and second differentiation matrices along the wall-normal coordinate, and $\boldsymbol{Z}$ is the zeros matrix.
No-slip conditions at the walls are applied to the linear operator.

The input matrix $\boldsymbol{B}$ is given by
\begin{equation}
	\boldsymbol{B} =
	\left[\begin{array}{ccc}
		\boldsymbol{I}	& \boldsymbol{Z}	& \boldsymbol{Z} \\
		\boldsymbol{Z}	& \boldsymbol{I}	& \boldsymbol{Z} \\
		\boldsymbol{Z}	& \boldsymbol{Z}	& \boldsymbol{I} \\
		\boldsymbol{Z}	& \boldsymbol{Z}	& \boldsymbol{Z}
	\end{array}\right] \mbox{,}
	\label{eq:actuation_matrix}
\end{equation}
\noindent with the rows corresponding to the wall positions for the three momentum equations set to zero to enforce the boundary conditions.

To restrict the observations to the pressure field, operator $\boldsymbol{C}$ is defined in this work as
\begin{equation}
	\boldsymbol{C} =
	\left[\begin{array}{cccc}
		\boldsymbol{Z}	& \boldsymbol{Z}	& \boldsymbol{Z}	& \boldsymbol{I}
	\end{array}\right] \mbox{.}
	\label{eq:observation_matrix}
\end{equation}

The linear operator can include an eddy viscosity model.
Following \citet{towne2020resolvent}, the linear operator $\boldsymbol{L}$ including the Cess eddy viscosity model \citep{cess1958survey} reads 
\begin{equation}
	\boldsymbol{L} =
	\left[\begin{array}{cccc}
		i \alpha \boldsymbol{\bar{U}} - i \omega \boldsymbol{I} - \boldsymbol{E} - \boldsymbol{J}	& - i \alpha {\nu_T}^{\prime} \boldsymbol{I} + \boldsymbol{\frac{d \bar{U}}{d y}}	& \boldsymbol{Z}	& i \alpha \boldsymbol{I} \\
		\boldsymbol{Z} & i \alpha \boldsymbol{\bar{U}} - i \omega \boldsymbol{I} - 2 \boldsymbol{E} - \boldsymbol{J}	& \boldsymbol{Z}	& \boldsymbol{\frac{d}{d y}} \\
		\boldsymbol{Z} & - i \beta {\nu_T}^{\prime} \boldsymbol{I}	& i \alpha \boldsymbol{\bar{U}} - i \omega \boldsymbol{I} - \boldsymbol{E} - \boldsymbol{J}	& i \beta \boldsymbol{I} \\
		i \alpha \boldsymbol{I}	& \boldsymbol{\frac{d}{d y}}	& i \beta \boldsymbol{I}	& \boldsymbol{Z}
	\end{array}\right] \mbox{,}
	\label{eq:eddy_viscosity_operator}
\end{equation}
\noindent where ${\nu_T}^{\prime} = \boldsymbol{\frac{d}{d y}} {\nu_T}$, $\boldsymbol{E} = {\nu_T}^{\prime} \boldsymbol{\frac{d}{d y}}$ and $\boldsymbol{J} = \frac{1}{Re_{\tau}} \frac{{\nu_T}}{\nu} \boldsymbol{\nabla^2}$.
The total eddy viscosity ${\nu_T}(y)$, obtained after the sum of the turbulent eddy viscosity ${\nu_t}(y)$ and the viscosity $\nu$, is modeled as
\begin{equation}
	frac{{\nu_T}}{\nu} = \frac{1}{2} \left\{1 + \frac{\kappa^2 {Re_{\tau}}^2}{9} \left[1 - \left(\eta - 1\right)^2\right]^2 \left[1 + 2 \left(\eta - 1\right)^2\right]^2 \left[1 - e^{\left(|\eta - 1| - 1\right) Re_{\tau} / A}\right]^2\right\}^{1/2} + \frac{1}{2} \mbox{,}
	\label{eq:eddy_viscosity_Cess}
\end{equation}
\noindent where the constants $\kappa$ and $A$ are constants given as 0.426 and 25.4, respectively \citep{pujals2009note}, and $\eta$ is the non-dimensional wall distance in outer units, i.e. $\eta = y/H$ and $y \in \left[0,2H\right]$ (i.e. $\eta \in \left[0,2\right]$).

The application of singular-value decomposition (SVD) on the resolvent operator accounting for the integration weights leads to
\begin{subeqnarray}
	{\boldsymbol{K}_{p}}^{1/2} \boldsymbol{H} {\boldsymbol{K}_{f}}^{-1/2} &=& \tilde{\boldsymbol{U}} \boldsymbol{\Sigma} \tilde{\boldsymbol{V}}^{\dagger} \mbox{,} \\
	\boldsymbol{V} &=& {\boldsymbol{K}_{f}}^{-1/2} \tilde{\boldsymbol{V}} \mbox{,} \\
	\boldsymbol{U} &=& \boldsymbol{R} \boldsymbol{B} \boldsymbol{V} \mbox{,}
	\label{eq:svd}
\end{subeqnarray}
\noindent where the columns of $\boldsymbol{U}$ and $\boldsymbol{V}$ form orthonormal bases, $\boldsymbol{\Sigma}$ is the a rectangular diagonal matrix of singular values and $\dagger$ denotes the Hermitian operator.
$\boldsymbol{U} = [\boldsymbol{U}_1~\boldsymbol{U}_2~\dots~\boldsymbol{U}_n]$ and $\boldsymbol{V} = [\boldsymbol{V}_1~\boldsymbol{V}_2~\dots~\boldsymbol{V}_n]$ can be interpreted as the optimal output (response or resolvent) and input (forcing) modes, respectively.
The gain matrix $\boldsymbol{\Sigma}$ is diagonal with real, positive and decreasing values, i.e. $\sigma_1 \geq \sigma_2 \geq \dots \geq \sigma_n$.
A harmonic forcing $\boldsymbol{\hat{f}} = \boldsymbol{V}_i$ leads to a flow response $\boldsymbol{\hat{q}} = \sigma_i \boldsymbol{U}_i$.
For $n = 1$, the pair $(\boldsymbol{U}_1,\boldsymbol{V}_1)$ provides the optimal, or most amplified, forcing and its associated response \citep{mckeon2010critical, hwang2010linear}.
When $\sigma_1 \gg \sigma_2$, i.e. there exists a large gain separation between the first two singular values, the first response mode tends to dominate the flow and the operator is denominated as low-rank.

Tilde notations in equation \ref{eq:svd} indicate quantities biased by the integration weights, with the Cholesky decomposition denoted by ${\boldsymbol{K}_{p}}^{1/2}$ and ${\boldsymbol{K}_{f}}^{1/2}$ regarding the pressure and forcing components, respectively.
$\boldsymbol{K}_{p}$ is a diagonal matrix containing Clenshaw-Curtis quadrature weights \citep{trefethen2000spectral} and $\boldsymbol{K}_{f}$ is defined as
\begin{equation}
	\boldsymbol{K}_{f} =
	\left[\begin{array}{ccc}
		\boldsymbol{K}_{p}	& \boldsymbol{Z}			& \boldsymbol{Z} \\
		\boldsymbol{Z}			& \boldsymbol{K}_{p}	& \boldsymbol{Z} \\
		\boldsymbol{Z}			& \boldsymbol{Z}			& \boldsymbol{K}_{p} 
	\end{array}\right] \mbox{.}
	\label{eq:spod_forcing_weights}
\end{equation}

Note that if one defines equation \ref{eq:svd}c as
\begin{equation}
	\boldsymbol{U} = \boldsymbol{C} \boldsymbol{R} \boldsymbol{B} \boldsymbol{V} = \boldsymbol{H} \boldsymbol{V} = {\boldsymbol{K}_{p}}^{-1/2} \tilde{\boldsymbol{U}} \mbox{,}
	\label{eq:response_mode}
\end{equation}
\noindent the resolvent modes would be restricted by the pressure fluctuation component, i.e. only pressure modes could be recovered.
Defining equation \ref{eq:svd}c as $\boldsymbol{U} = \boldsymbol{R} \boldsymbol{B} \boldsymbol{V}$ enables the recovering of the full-state modes, i.e. velocity and pressure fluctuation components.

\subsection{Two-point statistics (CSD)}
\label{sec:CSD}

Let us consider
\begin{subeqnarray}
	\boldsymbol{Q} &=& \left\langle \boldsymbol{\hat{q}} \boldsymbol{\hat{q}}^{\dagger} \right\rangle \mbox{,} \\
	\boldsymbol{F} &=& \left\langle \boldsymbol{\hat{f}} \boldsymbol{\hat{f}}^{\dagger} \right\rangle \mbox{,}
	\label{eq:state_forcing_CSD}
\end{subeqnarray}
\noindent where $\boldsymbol{Q}$ and $\boldsymbol{F}$ are the cross-spectral densities (CSDs), or the two-point statistics, of the state and forcing components as a function of $(\alpha,~\beta,~\omega)$ and symbol $\left\langle \dots \right\rangle$ denoting the expected-value operator.
Welch's method \citep{welch1967fft} can be employed to evaluate a CSD from a time series.

It is possible to obtain the flow response to an stochastic process employing the resolvent operator \citep{towne2018spectral, lesshafft2019resolvent, cavalieri2019wave}, i.e.
\begin{equation}
	\boldsymbol{S} = \boldsymbol{H} \boldsymbol{F} \boldsymbol{H}^{\dagger} \mbox{.}
	\label{eq:PqqPff}
\end{equation}
Moreover, the observation CSD can be obtained as 
\begin{equation}
	\boldsymbol{Y} = \boldsymbol{C} \boldsymbol{Q} \boldsymbol{C}^{\dagger} \mbox{.}
	\label{eq:observation_CSD}
\end{equation}

Expression \ref{eq:PqqPff} is the input-output relation between the forcing ($\boldsymbol{F}$) and response ($\boldsymbol{S}$) CSDs.
Knowledge of the full forcing statistics terms allows the recovery of the full response statistics.
If the forcing statistics terms are unknown, one option is to model it as white noise, i.e. $\boldsymbol{F} = \boldsymbol{I}$ and hence
\begin{equation}
	\boldsymbol{S}_I = \boldsymbol{H} \boldsymbol{H}^{\dagger} \mbox{.}
	\label{eq:resolvent_CSD}
\end{equation}
In practical applications, typically the forcing CSD terms are unknown and modelling them as white noise is an interesting possibility.
Yet, one may include an eddy viscosity model in the linear operator formulation to account for part of the forcing terms maintaining the forcing CSD as white noise, i.e.
\begin{equation}
	\boldsymbol{S}_{{\nu_T}} = \boldsymbol{H}_{{\nu_T}} {\boldsymbol{H}_{{\nu_T}}}^{\dagger} \mbox{,}
	\label{eq:resolvent_eddy_viscosity_CSD}
\end{equation}
\noindent where subscript ${\nu_T}$ indicates the use of the eddy viscosity model within the resolvent operator.

\subsection{SPOD}
\label{sec:spod_modeling}

Spectral proper orthogonal decomposition (SPOD) is a technique that decomposes a set of snapshots into an orthonormal, optimal basis in the frequency domain \citep{berkooz1993proper, taira2017modal, towne2018spectral}.
The obtained modes represent structures that maximize the mean square energy of the analyzed flow field for a given inner product \citep{picard2000pressure}.

The SPOD can be evaluated by taking the eigenvalue decomposition of the cross-spectral density (CSD) of the state (or forcing) for a triplet $(\alpha,~\beta,~\omega)$ and an appropriate weight or inner product.
In discrete form, the generalized problem can be written as
\begin{equation}
	\boldsymbol{D} \boldsymbol{W} \boldsymbol{\Psi} = \boldsymbol{\Psi} \boldsymbol{\Lambda} \mbox{,}
	\label{eq:spod}
\end{equation}
\noindent where $\boldsymbol{D}$ is the CSD of interest, e.g. state observations ($\boldsymbol{Y}$, equation \ref{eq:observation_CSD}) or forcing ($\boldsymbol{F}$, equation \ref{eq:state_forcing_CSD}b), $\boldsymbol{W}$ accounts for both the integration weights and the numerical quadrature on the discrete grid, $\boldsymbol{\Psi}$ denotes the SPOD modes and $\boldsymbol{\Lambda}$ corresponds to the eigenvalues.

The SPOD modes are an orthonormal basis that optimally represent the CSD of interest, in the present case the pressure fluctuations, for a given wavenumber-frequency combination \citep{towne2018spectral}.
The contribution of each SPOD mode (eigenfunction) to the CSD representation is given by its associated eigenvalue, ranked according to the energetic content, i.e. from the most to the least energetic, i.e. $\lambda_1 \geq \lambda_2 \geq \cdots \lambda_{n}$.

To extract the SPOD modes, we followed the procedures by \citet{towne2018spectral} and \citet{schmidt2019efficient}.
Therefore, we have not solved equation \ref{eq:spod} but an equivalent problem that contains the same non-zero eigenvalues, given by
\begin{subeqnarray}
	\boldsymbol{\hat{G}}^{\dagger} \boldsymbol{W} \boldsymbol{\hat{G}} \boldsymbol{\Theta} &=&  \boldsymbol{\Theta} \boldsymbol{\tilde{\Lambda}} \mbox{,} \\
	\boldsymbol{\tilde{\Psi}} &=& \boldsymbol{\hat{G}} \boldsymbol{\Theta} \boldsymbol{\tilde{\Lambda}}^{-1/2} \mbox{,}
	\label{eq:spod_true}
\end{subeqnarray}
\noindent where $\boldsymbol{\hat{G}}$ is a matrix containing the state (or forcing) Fourier transforms of each data block, $\boldsymbol{\Theta}$ denotes the eigenvectors, $\boldsymbol{\tilde{\Lambda}}$ stands for the eigenvalues and $\boldsymbol{\tilde{\Psi}}$ contains the SPOD modes.
The total number of non-zero eigenvalues is $N_b$, i.e. the same as the number of blocks employed in the Welch's method to obtain the CSDs.

To target the pressure SPOD, operator $\boldsymbol{W}$ is defined as
\begin{equation}
	\boldsymbol{W} =
	\left[\begin{array}{cccc}
		\boldsymbol{Z}	& \boldsymbol{Z}	& \boldsymbol{Z}	& \boldsymbol{Z} 			\\
		\boldsymbol{Z}	& \boldsymbol{Z}	& \boldsymbol{Z}	& \boldsymbol{Z} 			\\
		\boldsymbol{Z}	& \boldsymbol{Z}	& \boldsymbol{Z}	& \boldsymbol{Z}			\\
		\boldsymbol{Z}	& \boldsymbol{Z}	& \boldsymbol{Z}	& \boldsymbol{K}_{p}	\\
	\end{array}\right] \mbox{,}
	\label{eq:spod_pressure_weights}
\end{equation}
\noindent whereas $\boldsymbol{\hat{G}}$ is a matrix that is organized, for each frequency $\omega$ and wavenumber pair $(\alpha,~\beta)$, as containing the Fourier transforms of each state component and for each wall-normal distance $y$ in its lines, i.e. $N_y$ lines ($N_y$ being the mesh discretization in the wall-normal direction), and the columns are organized in terms of data blocks, i.e. $N_b$ columns.
Therefore,
\begin{equation}
	\boldsymbol{\hat{G}} = \left[\begin{array}{cccc} \boldsymbol{\hat{q}}_1	&	\boldsymbol{\hat{q}}_2	& \cdots	& \boldsymbol{\hat{q}}_{N_b} \end{array}\right] \mbox{,}
	\label{eq:G_matrix}
\end{equation}
\noindent for a triplet $\left(\alpha,~\beta,~\omega\right)$ and 
\begin{equation}
	\boldsymbol{\hat{q}} = \kappa \left[\begin{array}{cccc}\boldsymbol{\hat{u_x}} & \boldsymbol{\hat{u_y}} & \boldsymbol{\hat{u_z}} & \boldsymbol{\hat{p}} \end{array}\right]^{T} \mbox{,}
	\label{eq:q_matrix}
\end{equation}
\noindent where the state components are written in the frequency-wavenumber domain, i.e. for a triplet $\left(\alpha,~\beta,~\omega\right)$, and $\kappa$ is a constant defined as
\begin{equation}
	\kappa = \frac{1}{\sqrt{N_b \Delta f}} \mbox{,}
	\label{eq:kappa}
\end{equation}
\noindent where $\Delta f$ is the frequency discretization.

It is important to remark that resolvent and SPOD modes are identical when the forcing CSD is considered white in space, with the SPOD eigenvalues equal to the square of the resolvent modes \citep{towne2018spectral}.

\subsection{Numerical simulations}
\label{sec:dns}

\begin{figure}[!h]
	\centerline{\includegraphics[width=0.6\textwidth]{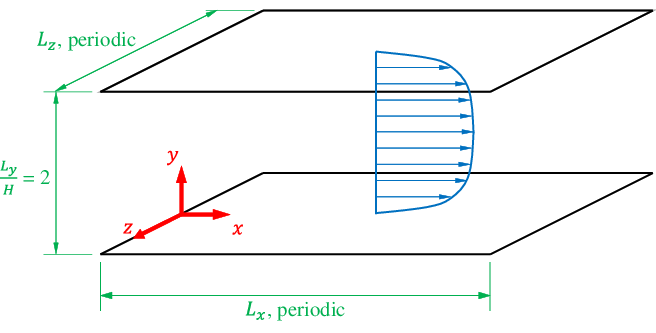}}
	\caption{Sketch of the channel flow geometry (black), coordinate system (red) and mean flow (blue), together with relevant dimensions (green).}
	\label{fig:channel_flow_sketch}
\end{figure}

The ChannelFlow pseudo-spectral code \citep{gibson2019channelflow} was employed to conduct the DNS of incompressible turbulent channel flow for friction Reynolds numbers of $Re_{\tau} \approx 180$ and 550.
In Cartesian coordinates, with half height $H$ used for normalization, the boxes dimensions are $4\pi \times 2 \times 2\pi$ in the streamwise ($L_x$), wall-normal ($L_y$) and spanwise ($L_z$) directions, respectively, for the $Re_{\tau} \approx 180$ simulation, and $2\pi \times 2 \times \pi$ for the $Re_{\tau} \approx 550$ case.
The wall-normal direction is defined within $y/H \in \left[0,~2\right]$.
Figure \ref{fig:channel_flow_sketch} shows a sketch of the flow geometry employed in the simulations.

For each simulation, the state and forcing components were stored as a function of time $t$, wall-normal distance $y$ and streamwise and spanwise wavenumbers, $\alpha$ and $\beta$, respectively, i.e. $\boldsymbol{q} = \boldsymbol{q}\left(\alpha,~y,~\beta,~t\right)$ and $\boldsymbol{f} = \boldsymbol{f}\left(\alpha,~y,~\beta,~t\right)$.
Chebyshev polynomials were employed for the discretization of the wall-normal direction ($y$), whereas the streamwise ($x$) and spanwise ($z$) directions were discretized using Fourier modes; the collocation points in the streamwise and spanwise directions were expanded by a 3/2 factor for de-aliasing purposes \citep{kim1987turbulence}.

Table \ref{tab:dns_parameters} shows the simulation main parameters, including the bulk Reynolds numbers $Re$ and the friction Reynolds numbers $Re_{\tau}$, the number of mesh points ($N_x$, $N_y$ and $N_z$) and mesh discretization ($\Delta x^{+}$, $\Delta z^{+}$, $\Delta {y_{min}}^{+}$ and $\Delta {y_{max}}^{+}$).
The first column of the table contains the canonical values for the friction Reynolds number and, between parenthesis, their actual value from the simulations.
The plus superscript indicates non-dimensional quantities using inner (viscous) scaling.
Table \ref{tab:dns_parameters} also includes information on the number of snapshots ($N_t$) and the time steps based on inner units ($\Delta t^{+}$), as well as  the block-size ($N_{fft}$) and number-of-blocks ($N_b$) employed to evaluate CSDs and PSDs.
The computation of power spectra, CSDs and PSDs of the flow state and forcing were performed using Welch's method \citep{welch1967fft} with a Hann window and blocks containing 75\% overlap.

\begin{table}[!h]
  \begin{center}
		\def~{\hphantom{0}}
		\caption{DNS parameters of the two channel flow cases addressed in the present study.}
		\begin{tabular}{c c c c c c c c c c c c c}
			$Re_{\tau}$	& $Re$  	& $N_x$	& $N_y$	& $N_z$	& $\Delta x^{+}$	& $\Delta z^{+}$	& $\Delta {y_{min}}^{+}$	& $\Delta {y_{max}}^{+}$	& $N_t$	& $\Delta t^{+}$	& $N_{fft}$	& $N_b$	\\ [3pt]
			\hline \\ [3pt]
			180 (179)		& 2,800		& 192		& 129		& 192		& 11.71						& 5.86						& 5.39 $\times$ 10$^{-2}$	& 4.39										& 2,399	& 5.72						& 64				& 146		\\
			550 (543)		& 10,000	& 384 	& 257		& 384		& 8.89						& 4.44						& 4.09 $\times$ 10$^{-2}$	& 6.66										& 3,000	& 2.95						& 64				& 184		\\
		\end{tabular}
		\label{tab:dns_parameters}
  \end{center}
\end{table}

\begin{figure}[!h]
	\centering
	\begin{subfigure}{0.48\textwidth}
		\includegraphics[width=\textwidth]{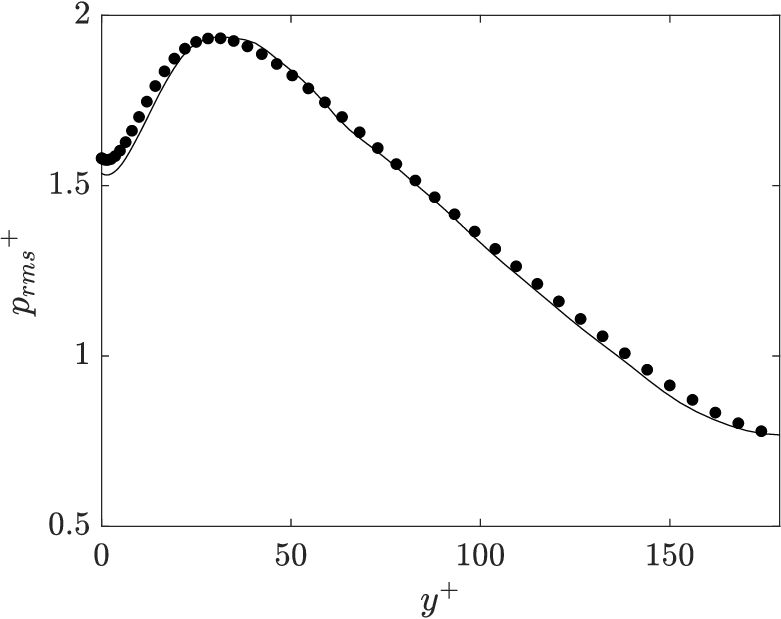}
		\caption{$Re_{\tau} \approx 180$}
		\label{fig:prms_validation_Retau180}
	\end{subfigure}
	\begin{subfigure}{0.48\textwidth}
		\includegraphics[width=\textwidth]{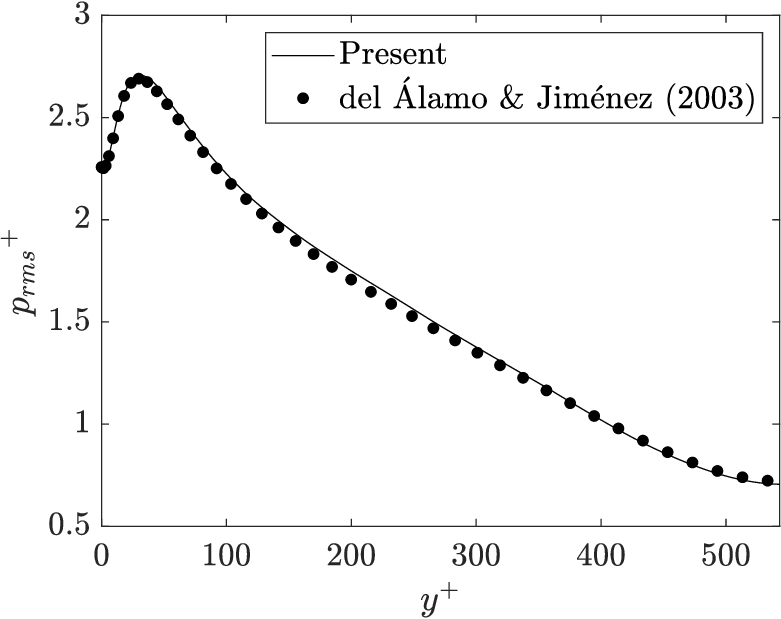}
		\caption{$Re_{\tau} \approx 550$}
		\label{fig:prms_validation_Retau550}
	\end{subfigure}
	\caption{Comparison between the pressure component rms for the present simulations (continuous lines) and those by \citet{delalamo2003spectra} (symbols).}
	\label{fig:prms_validation}
\end{figure}

The works by \citet{martini2020resolvent} and \citet{morra2021colour} present the DNS validation results for the mean velocity profile and root-mean-square (rms) values of velocity components; the former paper for $Re_{\tau} \approx 180$ and the later for $Re_{\tau} \approx 550$.
Literature results from earlier simulations \citep{delalamo2003spectra} were employed for validation purposes.
Pressure statistics validation results were not addressed by \citet{martini2020resolvent} and \citet{morra2021colour}.
Therefore, figure \ref{fig:prms_validation} displays the root-mean-square of the pressure component for the present database (continuous lines) and for the data obtained from \citet{delalamo2003spectra} (symbols), which show good agreement.

\begin{figure}[!h]
	\centering
	\begin{subfigure}{0.48\textwidth}
		\includegraphics[width=\textwidth]{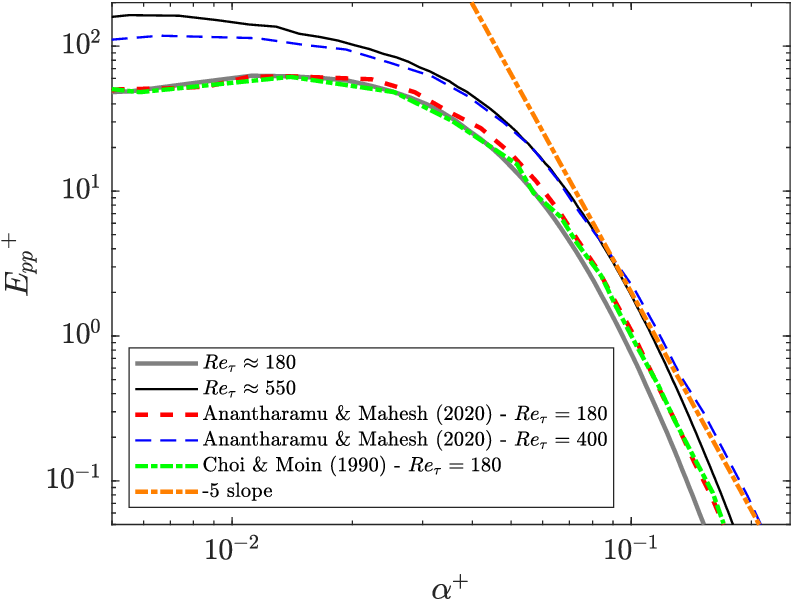}
		\label{fig:energy_spectra_1D_alpha}
	\end{subfigure}
	\begin{subfigure}{0.48\textwidth}
		\includegraphics[width=\textwidth]{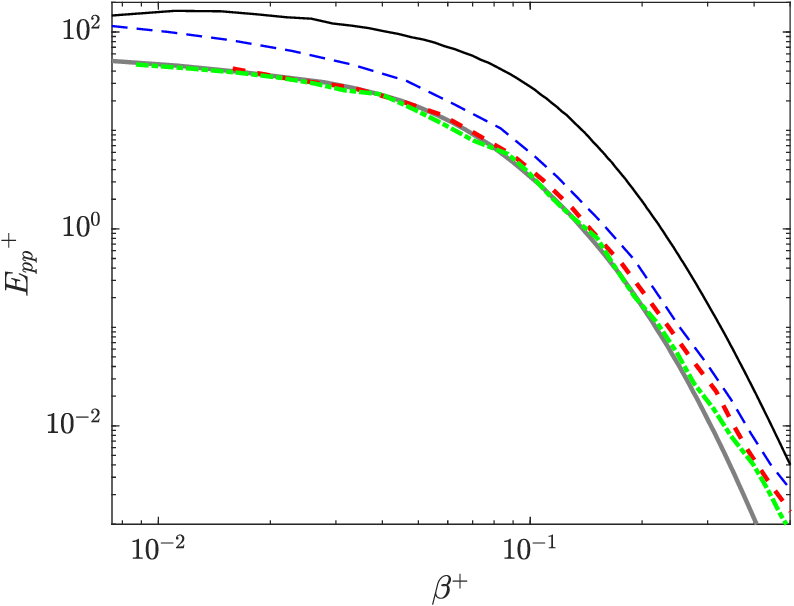}
		\label{fig:energy_spectra_1D_beta}
	\end{subfigure}	
	\caption{Comparison among wall-pressure fluctuation streamwise wavenumber spectra for the present $Re_{\tau} \approx 180$ (gray thick continuous line) and 550 (black thin continuous line) simulations and the results by \citet{anantharamu2020analysis} at $Re_{\tau} \approx 180$ (red thick dashed line) and 400 (blue thin dashed line), and by \citet{choi1990space} at $Re_{\tau} \approx 180$ (green thick dash-dotted line). Left frame: spectra as a function of streamwise wavenumber in inner units. Right frame: spectra as a function of spanwise wavenumber in inner units. A -5 slope curve (orange thick dotted line) is also plotted in the left frame.}
	\label{fig:energy_spectra_1D}
\end{figure}

To further validate the DNS databases, figure \ref{fig:energy_spectra_1D} displays one-sided wall-pressure spectra scaled in inner units as a function of streamwise (left frame) and spanwise (right frame) wavenumbers.
Comparisons among the results of the present simulations with others obtained in the literature \citep{choi1990space, anantharamu2020analysis} are displayed in the figure.
Note the $Re_{\tau} \approx 550$ database has a higher friction Reynolds number than the one used by \citet{anantharamu2020analysis}, which is 400.
Regarding the streamwise wavenumber spectra (left frame), good agreement is achieved for the energy containing part of the spectrum, at low wavenumbers.
Differences with respect to the $Re_{\tau} = 400$ case in \citet{anantharamu2020analysis} may be attributed to the different Reynolds numbers of the simulations.
The high wavenumber content of the present simulation has a faster decay than earlier results.
This may be related to the use of the rotational form of the nonlinear terms in the present simulation, as different choices for the nonlinear terms may affect results \citep{zang1991rotation}.
As the present study focuses on larger pressure structures, the agreement at lower wavenumbers and rms profiles was considered sufficient.

The present turbulent channel flow simulations have statistical symmetry around the central plane.
We take advantage of this by decomposing fluctuations into even and odd parts around the central plane, as in \citet{abreu2020resolvent}.
In this study we considered the odd part of pressure fluctuations, with the corresponding even or odd parts for the other state and forcing components.

\section{Wavenumber and frequency spectra of pressure fluctuations}
\label{sec:energy_spectra}

We start by evaluating wavenumber and frequency spectra for the pressure in order to select the most energetic structures for further analysis.
The procedure will follow the approach of \citet{morra2019relevance}, this time for pressure fluctuations.
Spectra without and with premultiplication will be used, since, as will be shown, energetic spanwise-coherent pressure structures appear in the spectrum without premultiplication.
A detailed analysis of the $Re_{\tau} \approx 550$ channel flow will be addressed in the following, whereas complementary results for the $Re_{\tau} \approx 180$ simulation are displayed in appendix \ref{app:Retau180_results}.

\begin{figure}[!h]
	\centerline{\includegraphics[width=\textwidth]{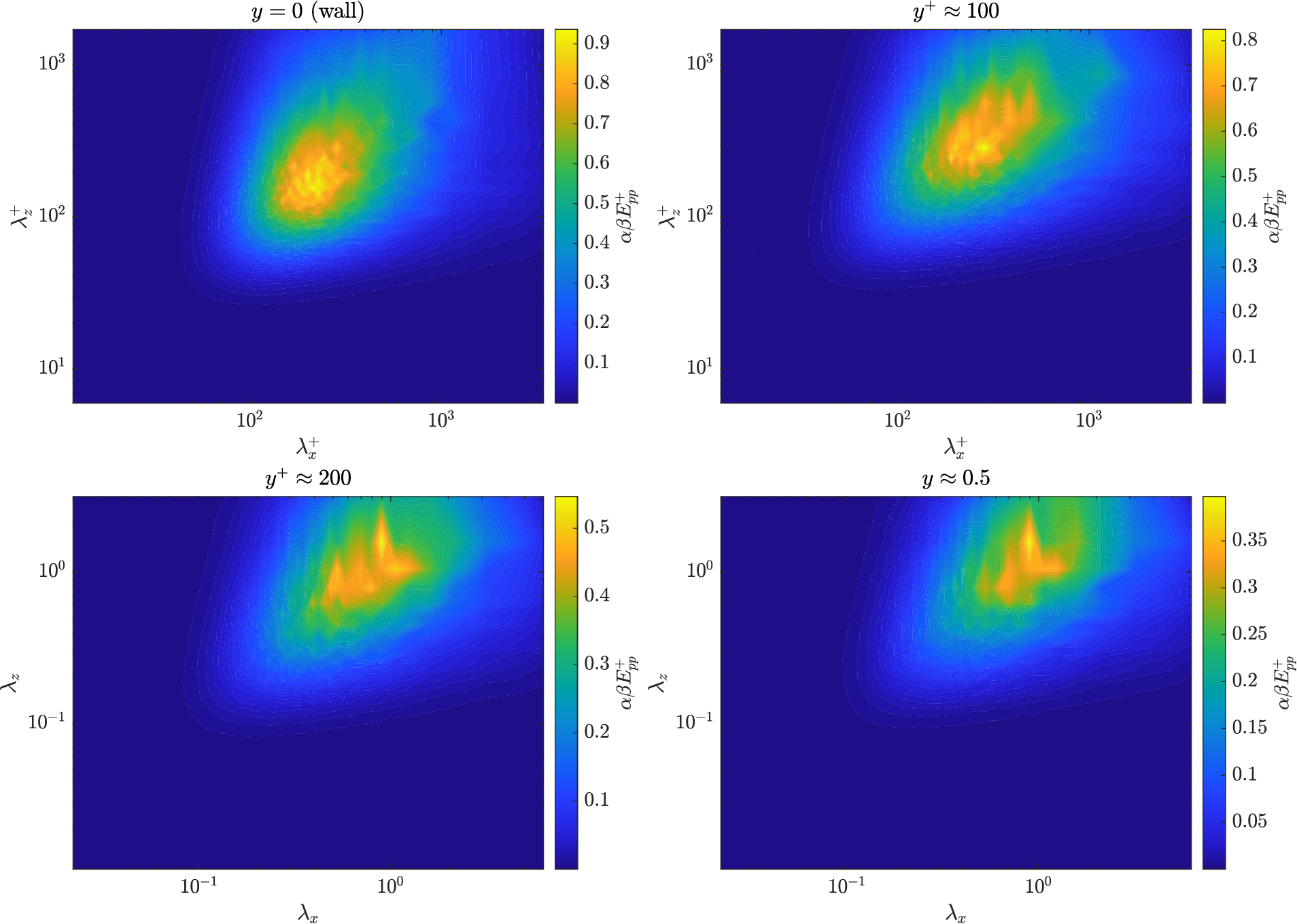}}
	\caption{Premultiplied energy spectra for $Re_{\tau} \approx 550$ simulation.}
	\label{fig:pmspectra_yp15_y05_Retau550}
\end{figure}

\begin{figure}[!h]
	\centerline{\includegraphics[width=\textwidth]{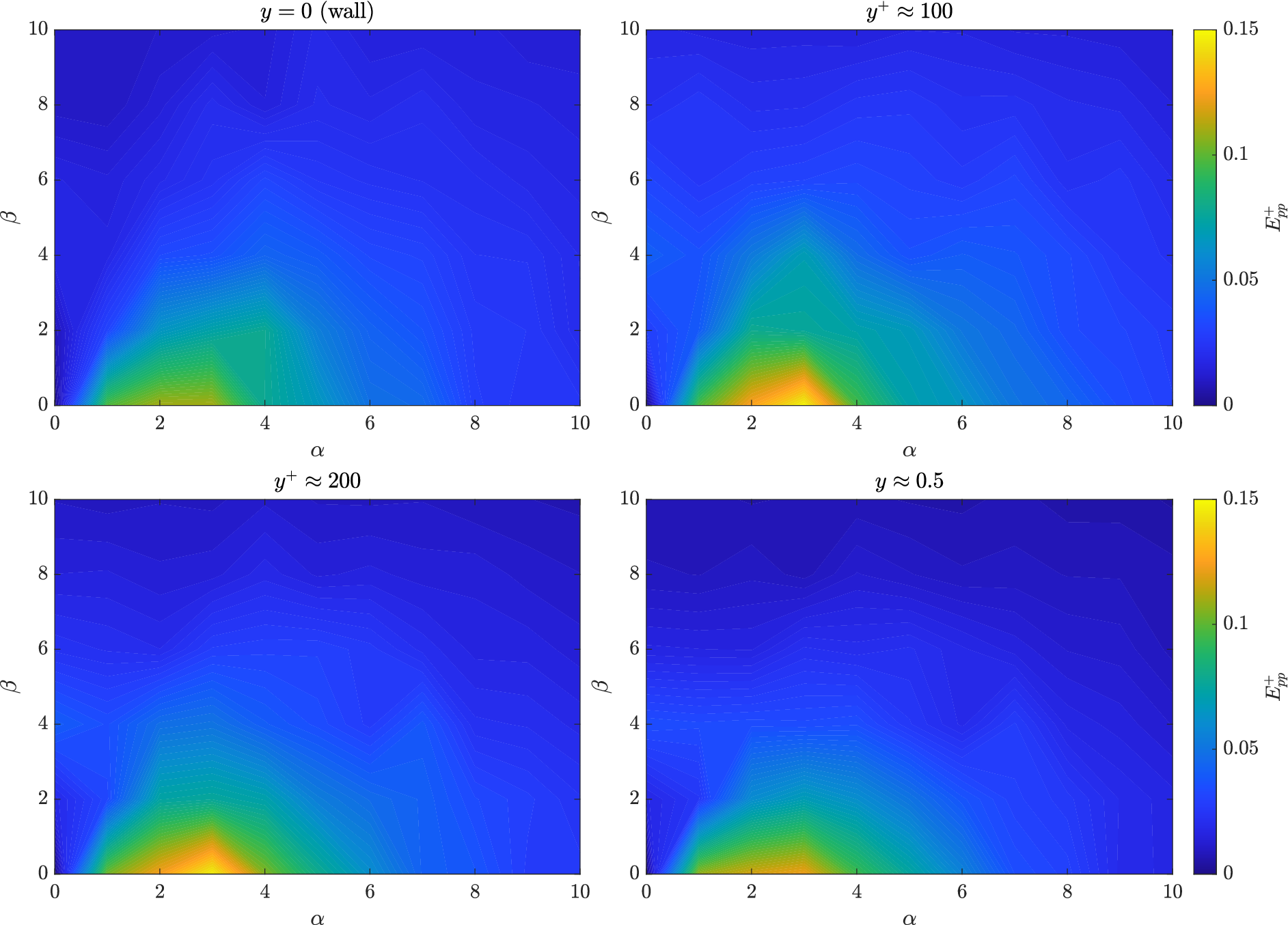}}
	\caption{Energy spectra for $Re_{\tau} \approx 550$ simulation.}
	\label{fig:spectra_y0_y05_Retau550}
\end{figure}

The pressure component spectra at the channel wall ($y = 0$) and at $y^+ \approx$ 100, 200 and $y \approx 0.5$ (a height characteristic of large-scale structures \citep{delalamo2003spectra}), are shown in figures \ref{fig:pmspectra_yp15_y05_Retau550} and \ref{fig:spectra_y0_y05_Retau550} for the pressure power spectra with and without premultiplication, respectively.
Minor differences between wall ($y = 0$) and buffer-layer spectra ($y^+ \approx 15$, characteristic of near-wall structures), were observed and hence we decided to show only the results on the channel wall.
The wall premultiplied spectrum peaks at $({\lambda_x}^+,~{\lambda_z}^+) \approx (227,~155)$ and the ${\lambda_x}^+$ peak value agrees with previous studies \citep{panton2017correlation, anantharamu2020analysis}.
On the other hand, the $y \approx 0.5$ plane premultiplied spectrum peaks at $(\lambda_x,~\lambda_z) \approx (0.90,~1.57)$.
The premultiplied spectra at $y^+ \approx 100$ resembles the one at $y = 0$, although the energy peak moved to slightly higher wavelengths, whereas the spectra at $y^+ \approx 200$ resembles the one at $y \approx 0.5$ (equivalent to $y^+ \approx 270$).
Note that on the wall ($y = 0$), peak spectral wavenumbers denote structures that are streamwise elongated, i.e. $\lambda_x > \lambda_z$, whereas far from the wall (e.g. $y \approx 0.5$), the more energetic structures are spanwise elongated, i.e. $\lambda_z > \lambda_x$.

The pressure spectra without premultiplication as a function of streamwise and spanwise wavenumbers shown in figure \ref{fig:spectra_y0_y05_Retau550} peaks at $(\alpha,~\beta) = (3,~0)$ for all the four studied planes, suggesting the presence of large spanwise elongated structures extending throughout the channel.
This peak wavenumber matches the results of \citet{yang2022wavenumber}.
\citet{abreu2021spanwise} and \citet{pozuelo2023widest} verified that such spanwise elongated structures ($\beta = 0$) for velocity fluctuations are not an artifact of small domains.
Spanwise-elongated pressure structures are potentially relevant for problems involving fluid-structure interaction \citep{kim2014space} and sound radiation \citep{sano2019trailing, demange2023resolvent}.
Hereinafter, as a convention, we will refer to the near-wall structures in inner (or wall) units, whereas large-scale structures will be dealt with using outer units.
We will also focus on the near-wall ($y = 0$), large-scale ($y \approx 0.5$) and the spanwise-coherent structures.

\begin{figure}[!h]
	\centerline{\includegraphics[width=\textwidth]{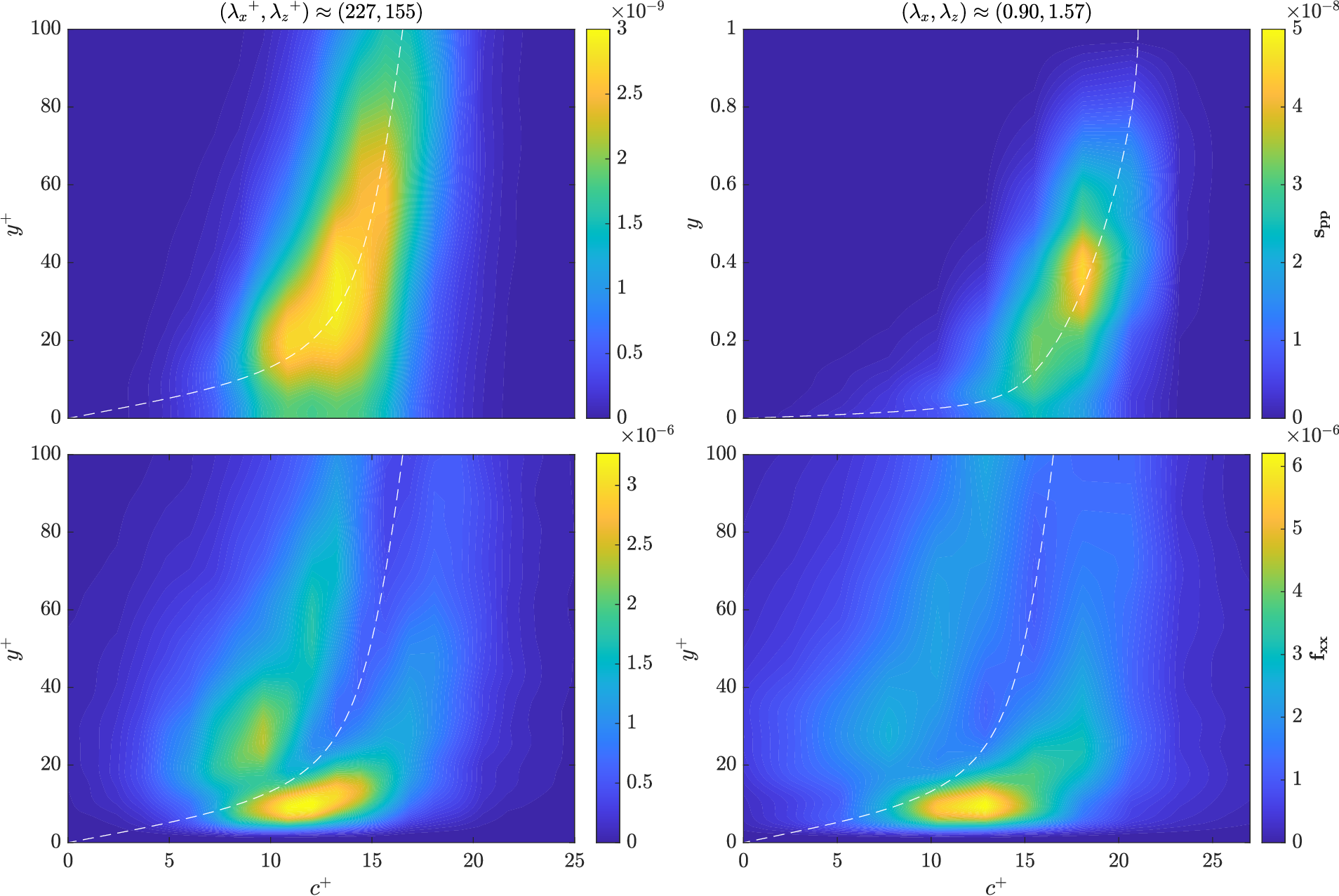}}
	\caption{PSDs for $Re_{\tau} \approx 550$ simulation. Left frames: near-wall structures. Right frames: large-scale structures. Top frames: pressure component ($\boldsymbol{s}_{pp}$). Bottom frames: streamwise forcing component ($\boldsymbol{f}_{xx}$). Dashed lines denote the critical layer. Note the vertical axis scaling of the top-left frame is in outer units ($y$), whereas the other frames retained inner units ($y^+$).}
	\label{fig:psd_y0_y05_Retau550}
\end{figure}

To determine frequencies of interest for the peak wavenumbers, similar to what was done by \citet{morra2019relevance}, the pressure PSD (power spectrum density) for near-wall and large-scale structures are exhibited in the top frames of figure \ref{fig:psd_y0_y05_Retau550} as a function of wall-normal distance and phase speed.
In the figure, the PSD is plotted as a function of the wall-normal distance and phase speed $c^+ = \omega^+/\alpha^+$, where $\omega^+$ is the angular frequency scaled in inner units.
The top and bottom frames correspond to the PSD of the pressure and streamwise forcing components, respectively.
The figure also shows the critical layer ($U^+ = c^+$) in dashed lines.
The pressure component PSDs peak at $c^+ \approx 13.5$ ($\omega^+ \approx 0.37$) with support at $y^+ \approx 30$ and $c^+ \approx 18$ ($\omega \approx 6.87$) with support at $y \approx 0.38$ for the near-wall and large-scale structures, respectively.
The $\omega^+$ peak value is in agreement with previous results \citep{hu2006wall, anantharamu2020analysis}.
For the forcing streamwise component, the PSDs peak at $c^+ \approx 12$ and 13 for the near-wall and large-scale structures, respectively, both with support at $y^+ \approx 10$.
Such near-wall nature of the forcing peak, even for large-scale structures, was reported by \citet{morra2021colour}.

\begin{figure}[!h]
	\centerline{\includegraphics[width=\textwidth]{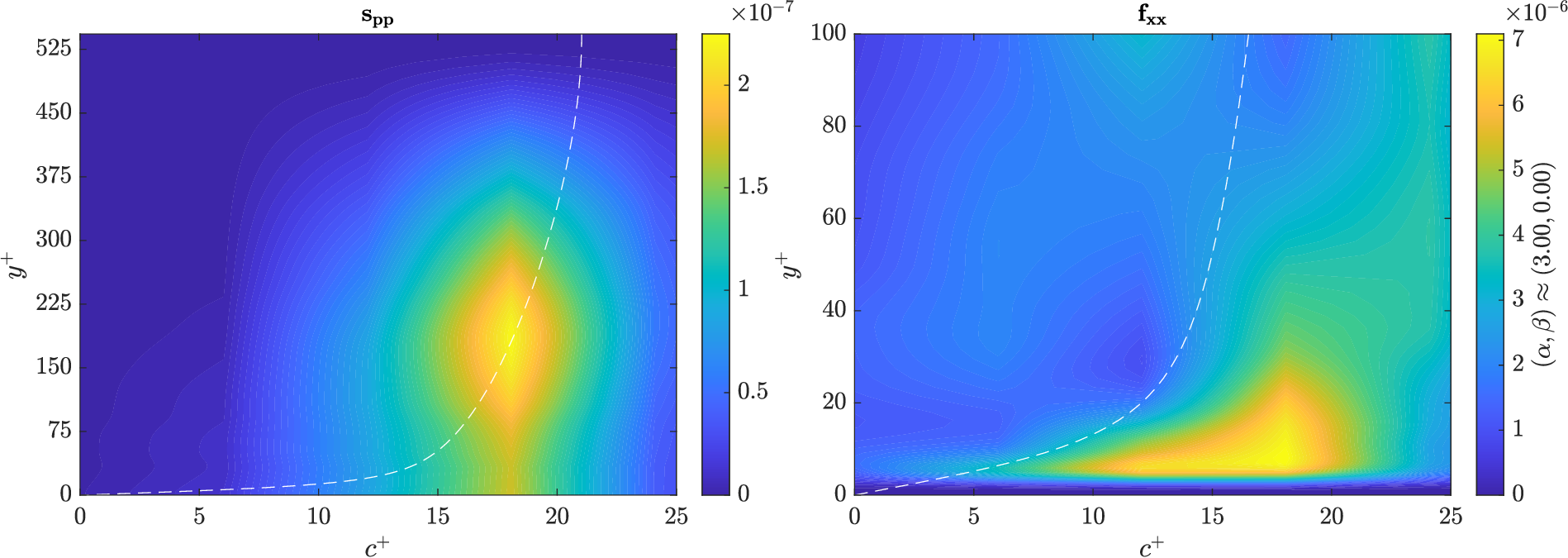}}
	\caption{PSDs for $Re_{\tau} \approx 550$ simulation and spanwise-coherent structure, i.e. $(\alpha,~\beta) = (3,~0)$. Right frame: pressure component ($\boldsymbol{s}_{pp}$). Left frame: streamwise forcing component ($\boldsymbol{f}_{xx}$). Dashed lines denote the critical layer.}
	\label{fig:psd_PSP_Retau550}
\end{figure}

The PSDs are shown in figure \ref{fig:psd_PSP_Retau550} for the pressure (right frame) and streamwise forcing (left frame) components for the spanwise-coherent structures with $(\alpha,~\beta) = (3,~0)$.
Both pressure and forcing components PSD peak at $\omega \approx 2.95$, the first with support at $y^+ \approx 183$, following the critical layer, and the latter with support at $y^+ \approx 7$.
Hereinafter, the power spectra without premultiplication peak will be refereed to as PSP.

Table \ref{tab:peak_values} summarizes the results of the preceding analysis of spectra for both Reynolds numbers.
The spectral peaks are given in terms of spanwise and streamwise wavenumbers and wavelengths (in inner and outer units), phase speeds, angular frequencies (in inner and outer units) and wall-normal distance of the peak (in inner and outer units), for both Reynolds numbers.c
Results for the pressure and streamwise forcing spectra are shown in the table, and when the peak location of the latter is different from the former ones, such values are provided between parenthesis.
Note that it is also possible to recover the temporal wavelengths through the relationships $\lambda_t = 2 \pi/\omega$ and ${\lambda_t}^+ = 2 \pi/{\omega^+}$, in outer and inner units, respectively.
The near-wall structures for both Reynolds numbers show similar peak location when considering inner scaling, i.e. $({\lambda_x}^+,~{\lambda_z}^+) \approx (200,~160)$ and $(c^{+},~\omega^+) \approx (13.5,~0.4)$, with support at $y^+ \approx 35$, whereas for the streamwise forcing component, one has $(c^{+},~\omega^+) \approx (12,~0.35)$.
Such scaling must be taken with care, since only two moderate Reynolds numbers are considered in the present study.
For completeness, results for the wall-normal distance of $y^+ \approx 15$ are also provided in the table and it is verified that the spectral characteristics at this distance are very similar to those obtained at the wall ($y = 0$).
For the large-scale structures ($y \approx 0.5$), no obvious scaling with either inner or outer units is observed.
Regarding the spectra without premultiplication, large structures that are infinitely long in the spanwise direction were found for both Reynolds numbers.
Analyzing the forcing spectra, such spanwise-coherent structures have at $y^+ \approx 7$ for both Reynolds numbers, whereas the pressure spectra show support at $y^+ \approx 44$ and 183 for $Re_\tau \approx 180$ and 550, respectively.

\begin{table}[!h]
  \begin{center}
		\def~{\hphantom{0}}
		\caption{Spectral peaks for wall, near-wall, large-scale and PSP structures for the two Reynolds numbers studied. Notation \emph{w/ pm} indicates spectra with premultiplication, whereas \emph{w/o pm} pm indicates spectra without premultiplication. When the streamwise forcing component peak location differs from those of the pressure component, the values of the forcing peak are given between parenthesis. ${y_{peak}}^+$ and $y_{peak}$ denote the PSD peak position.}
		\resizebox{\textwidth}{!}{
		\begin{tabular}{c c c c c c c c c c c c}
																															& ${\lambda_x}^+$	& ${\lambda_z}^+$	& $\lambda_x$	& $\lambda_z$	& $\alpha$	& $\beta$	& $c^{+}$		& $\omega^+$	& $\omega$			& ${y_{peak}}^+$		& $y_{peak}$					\\ [3pt]
			\hline \\ [3pt]
			\emph{w/ pm}, $y = 0$, $Re_{\tau} \approx 180$					& 204							& 161							& 1.14				& 0.90				& 5.5				& 7				& 14 (12.5)	& 0.43 (0.38)	& 4.91 (4.32)		& 38 (10)	& 0.21 (0.06)	\\
			\emph{w/ pm}, $y^+ \approx 15$, $Re_{\tau} \approx 180$	& 225							& 125							& 1.26				& 0.70				& 5					& 9				& 14 (12.5)	& 0.40 (0.34)	& 4.52 (3.93)		& 33 (10)	& 0.17 (0.06)	\\
			\emph{w/ pm}, $y \approx 0.5$, $Re_{\tau} \approx 180$	& 250							& 375							& 1.40				& 2.09				& 4.5				& 3				& 16 (12.5)	& 0.39 (0.31)	& 4.52 (3.53)		& 76 (10)	& 0.42 (0.06)	\\
			\emph{w/o pm}, $Re_{\tau} \approx 180$									& 749							& $\infty$				& 12.57				& $\infty$		& 0.5				& 0				& 12.5			& 0.03				& 0.39					& 44 (6.5)& 0.24 (0.04)	\\ [3pt]
			
			\emph{w/ pm}, $y = 0$, $Re_{\tau} \approx 550$					& 227							& 155							& 0.42				& 0.29				& 15				& 22			& 13.5 (12)	& 0.37 (0.33)	& 10.8	(9.82)	& 30 (10)	& 0.05 (0.02)	\\
			\emph{w/ pm}, $y^+ \approx 15$, $Re_{\tau} \approx 550$	& 227							& 131							& 0.42				& 0.24				& 15				& 26			& 12 (13)		& 0.33 (0.37)	& 9.82 (10.80)	& 20 (10)	& 0.04 (0.02)	\\		
			\emph{w/ pm}, $y \approx 0.5$, $Re_{\tau} \approx 550$	& 490							& 850							& 0.90				& 1.57				& 7					& 4				& 18 (13)		& 0.23 (0.17)	& 6.87 (4.91)		& 209 (10)& 0.38 (0.02)	\\
			\emph{w/o pm}, $Re_{\tau} \approx 550$									& 1137						& $\infty$				& 2.09				& $\infty$		& 3					& 0				& 18				& 0.10				& 2.95					& 183 (7)	& 0.34 (0.01)	\\
		\end{tabular}
		}
		\label{tab:peak_values}
  \end{center}
\end{table}

\section{Comparison between SPOD and resolvent modes}
\label{sec:spod_resolvent_analysis}

\subsection{Dominant structures}
\label{sec:dominant_structures}

The dominant coherent structures described in the previous section are further explored using SPOD and resolvent analysis for the $Re_{\tau} \approx 550$ simulation.
Appendix \ref{app:Retau180_results} displays complementary results for the $Re_{\tau} \approx 180$ simulation.
Regarding the resolvent analysis, results without and with the eddy viscosity model are addressed.

\begin{figure}[!h]
	\centerline{\includegraphics[width=\textwidth]{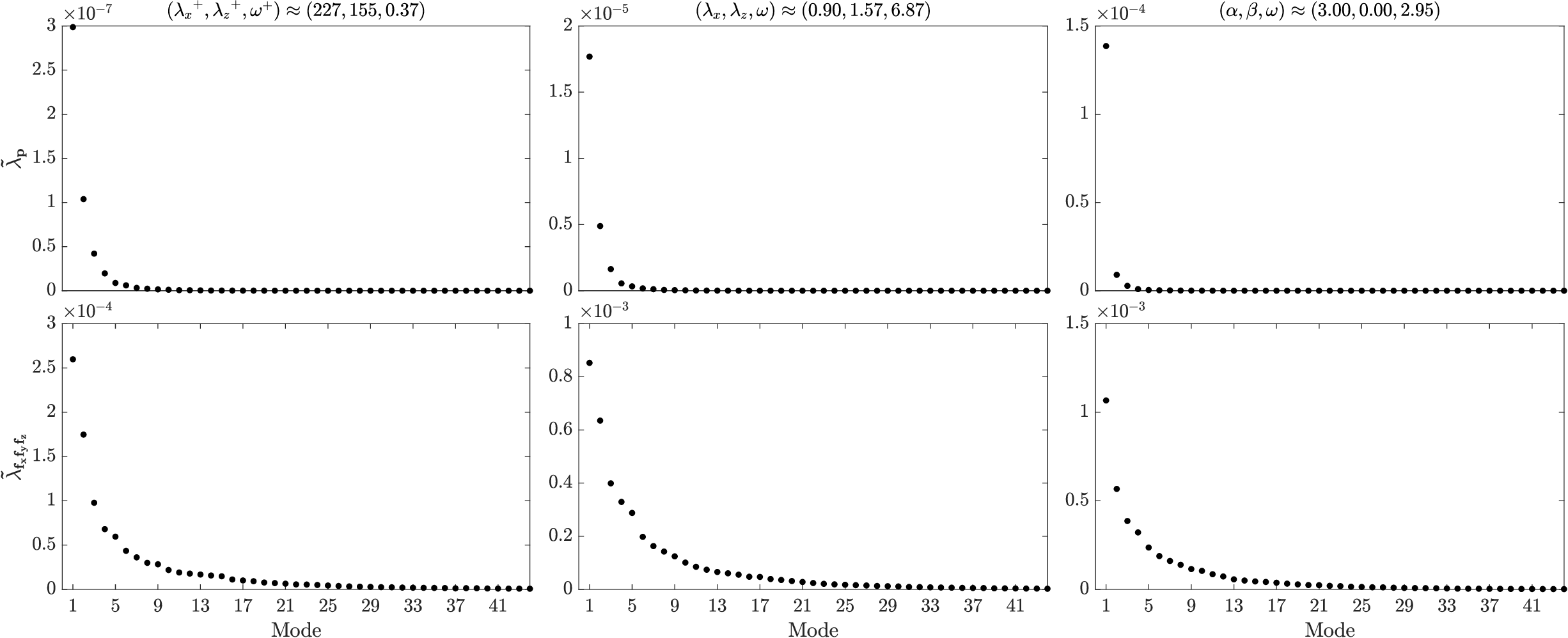}}
	\caption{SPOD eigenvalues employing the pressure (top frames) and forcing (bottom frames) modes for the $Re_{\tau} \approx 550$ simulation. Frames from left to right: near-wall, large-scale and spanwise-coherent structures, respectively.}
	\label{fig:spod_p_fxfyfz_eigenvalues_Retau550}
\end{figure}

Figure \ref{fig:spod_p_fxfyfz_eigenvalues_Retau550} exhibit the SPOD eigenvalues evaluated for the pressure (top frames) and forcing (bottom frames) components.
Near-wall, large-scale and spanwise-coherent structures are displayed in the left, middle and right frames, respectively.
The first SPOD eigenvalue correspond to approximately 61\%, 69\% and 91\% of the total energy for the near-wall, large scale and spanwise-coherent structures, respectively, when considering the pressure component.
\citet{abreu2020resolvent} report a first eigenvalue corresponding to about 68\% and 84\% of the total energy for the near-wall and large-scale structures, respectively.
Note that the near-wall and large-scale structures obtained by \citet{abreu2020resolvent} peak at other spatial and temporal wavelengths, since they are targeting the streamwise velocity component and consider the kinetic energy weighting to evaluate the SPOD modes.
If the forcing components are considered instead, the first SPOD eigenvalues decay to approximately 25\%, 21\% and 27\% of the total energy for the respective structures.
Taking into account the second forcing SPOD eigenvalue, raises the percentile contribution to approximately 43\%, 37\% and 41\%, respectively.
However, the leading pressure SPOD mode remains strongly dominant for the three considered structures.

\begin{figure}[!h]
	\centerline{\includegraphics[width=\textwidth]{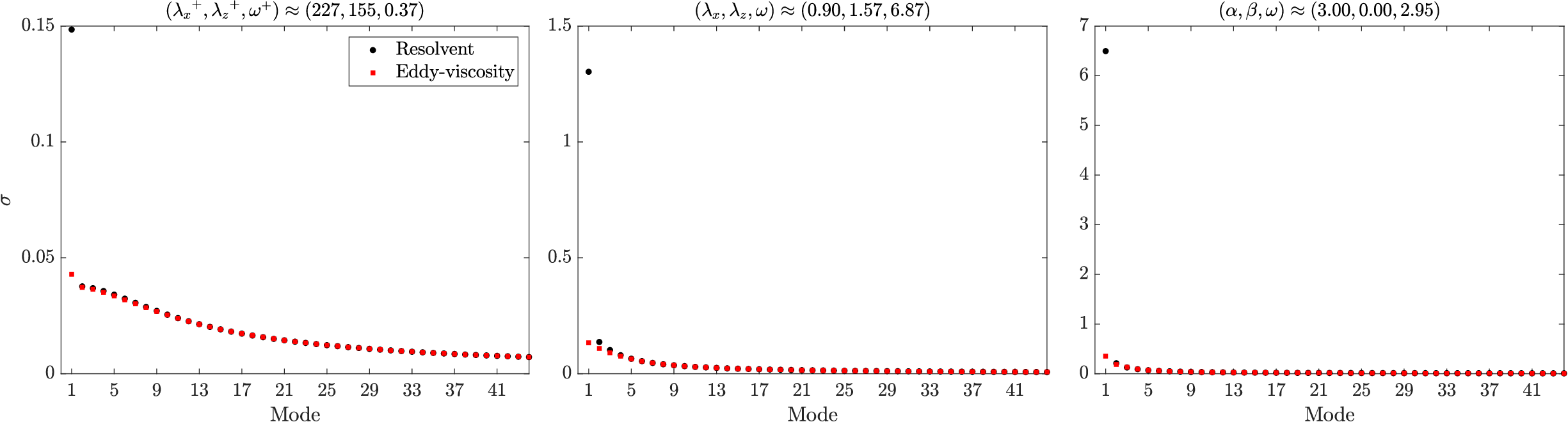}}
	\caption{Resolvent gains employing the pressure weighing for the $Re_{\tau} \approx 550$ simulation. Black circles: linear operator without eddy viscosity model. Red squares: linear operator with eddy viscosity model. See the comments in the caption of figure \ref{fig:spod_p_fxfyfz_eigenvalues_Retau550}.}
	\label{fig:resolvent_p_gains_Retau550}
\end{figure}

Figure \ref{fig:resolvent_p_gains_Retau550} displays the resolvent and eddy viscosity resolvent gains, where black circle markers indicate the former and red square markers account for the latter.
Frames, from left to right, denote the near-wall, large-scale and spanwise-coherent structures, respectively.
The ratio between the first and second resolvent gains, i.e. $\sigma_1/\sigma_2$, is of approximately 3.94, 9.52 and 30.85 for the near-wall, large scale and spanwise-coherent structures, respectively, i.e. a low-rank behavior is observed.
Regarding the eddy viscosity resolvent gains, such low-rank behavior is not present, with suboptimal gains close to the optimal ones.

\begin{figure}[!h]
	\centerline{\includegraphics[width=\textwidth]{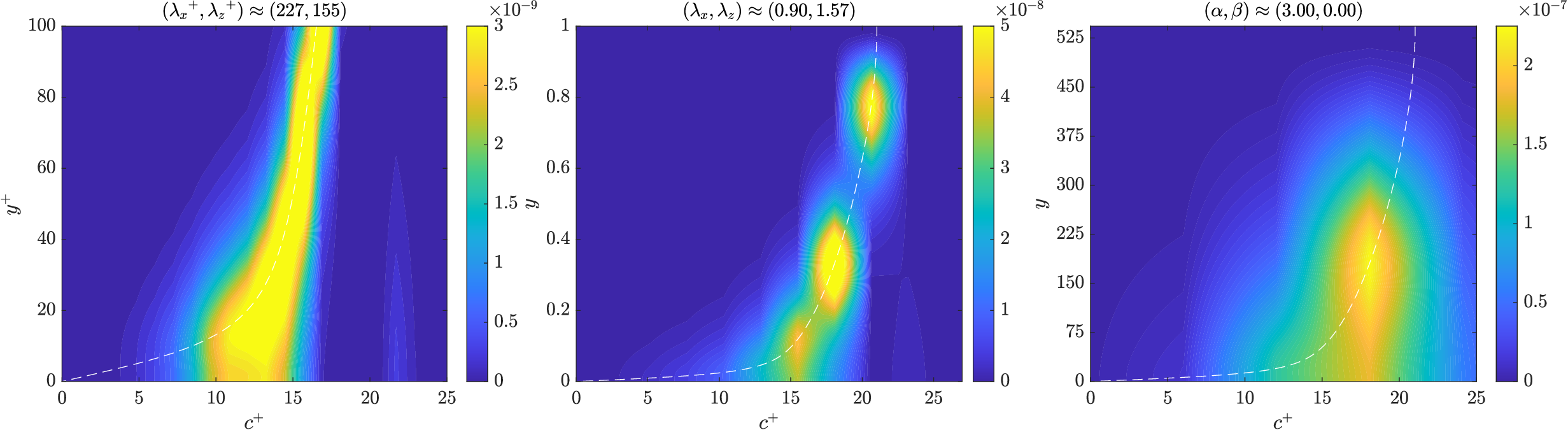}}
	\caption{Pressure component PSD low-rank resolvent reconstruction (1st mode) for the $Re_{\tau} \approx 550$ simulation. Frames from left to right: near-wall, large-scale and spanwise-coherent structures, respectively.}
	\label{fig:resolvent_PSD_reconstruction_Retau550}
\end{figure}

Low-rank reconstructions of the pressure PSD ($\boldsymbol{P}_{pp}^{(est)}$) for the near-wall and large-scale structures are displayed in figure \ref{fig:resolvent_PSD_reconstruction_Retau550} using only the first resolvent/response mode, i.e.
\begin{equation}
	\boldsymbol{P}_{pp}^{(est)} = a_{corr} \boldsymbol{U}_1 {\sigma_1}^2 {\boldsymbol{U}_1}^\dagger \mbox{,}
	\label{eq:PSD_rec}
\end{equation}
\noindent where $a_{corr}$ is the amplitude correction factor, chose to match the DNS ($\boldsymbol{P}_{pp}^{(DNS)}$, figures \ref{fig:psd_y0_y05_Retau550} and \ref{fig:psd_PSP_Retau550}) at each frequency $\omega$ as
\begin{equation}
	a_{corr}(\omega) = \frac{\int_{y/H=0}^{2} \boldsymbol{P}_{pp}^{(DNS)}(\alpha,y,y,\beta,\omega) dy}{\int_{y/H=0}^{2}\boldsymbol{U}_1 {\sigma_1}^2 {\boldsymbol{U}_1}^\dagger dy} \mbox{,}
	\label{eq:a_corr}
\end{equation}
\noindent in the same fashion as \citet{morra2019relevance}.

In comparison with the top frames of figure \ref{fig:psd_y0_y05_Retau550}, it is seen that the near-wall and large-scale structures are well represented by the first mode.
This feature is somewhat expected since a gain separation of approximately one order of magnitude is observed for such structure (figure \ref{fig:resolvent_p_gains_Retau550}).
However, the predictions from resolvent analysis without eddy viscosity tend to be more concentrated around the critical layer $c = U$ than the trends from DNS data \citep{morra2019relevance}; this feature is observed in the plots.
Regarding the spanwise-coherent structure, an even closer match with the DNS PSD is obtained, peaking at $(c^+,~y^+) \approx (18,~180)$; again, the gain separation is significant, which leads to a clearly dominant mechanism for spanwise-coherent structures.

\begin{figure}[!h]
	\centerline{\includegraphics[width=\textwidth]{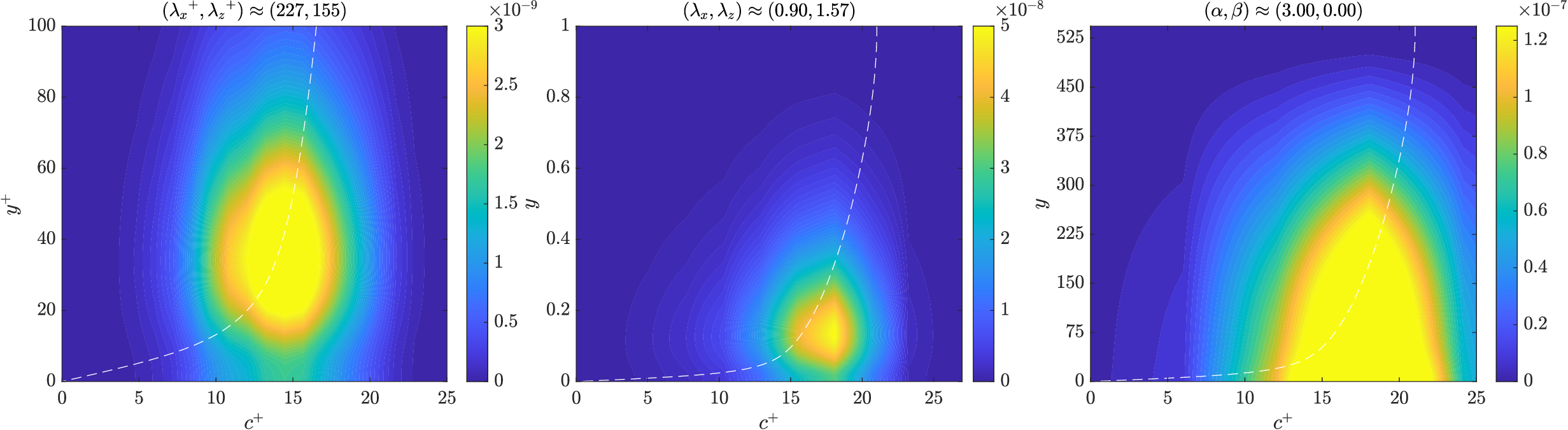}}
	\caption{Pressure component PSD low-rank eddy viscosity resolvent reconstruction (1st mode) for the $Re_{\tau} \approx 550$ simulation. See the comments in the caption of figure \ref{fig:resolvent_PSD_reconstruction_Retau550}.}
	\label{fig:resolventEV_PSD_reconstruction_Retau550}
\end{figure}

Figure \ref{fig:resolventEV_PSD_reconstruction_Retau550} shows the low-rank reconstructions now using the eddy viscosity resolvent, i.e.
\begin{equation}
	\boldsymbol{P}^{\nu_T (est)}_{pp} = a^{\nu_T}_{corr} \boldsymbol{U}_1^{\nu_T} {\sigma_1^{\nu_T}}^2 {\boldsymbol{U}_1^{\nu_T}}^\dagger \mbox{,}
	\label{eq:PSDeddy_rec}
\end{equation}
\noindent with
\begin{equation}
	a^{\nu_T}_{corr}(\omega) = \frac{\int_{y/H=0}^{2} \boldsymbol{P}_{pp}^{(DNS)}(\alpha,y,y,\beta,\omega) dy}{\int_{y/H=0}^{2} \boldsymbol{U}_1^{\nu_T} {\sigma_1^{\nu_T}}^2 {\boldsymbol{U}_1^{\nu_T}}^\dagger dy} \mbox{.}
	\label{eq:a_corr_nuT}
\end{equation}
Although this model does not show a low-rank behavior (figure \ref{fig:resolvent_p_gains_Retau550}), it captured the energy peak position for the near-wall structures, at $(c^+,~y^+) \approx (13.5,~30)$.
In contrast, the eddy viscosity resolvent does not well represent the energy peak position for the large-scale structure, which shows a peak at $(c^+,~y) \approx (16,~0.1)$, and for the spanwise-coherent structure, which only the phase speed peak is correct at $c^+ \approx 18$.
Besides the mismatch in pressure PSD peaks, the distributions along $y$ are not well represented by the low-rank reconstruction.

\begin{figure}[!h]
	\centerline{\includegraphics[width=\textwidth]{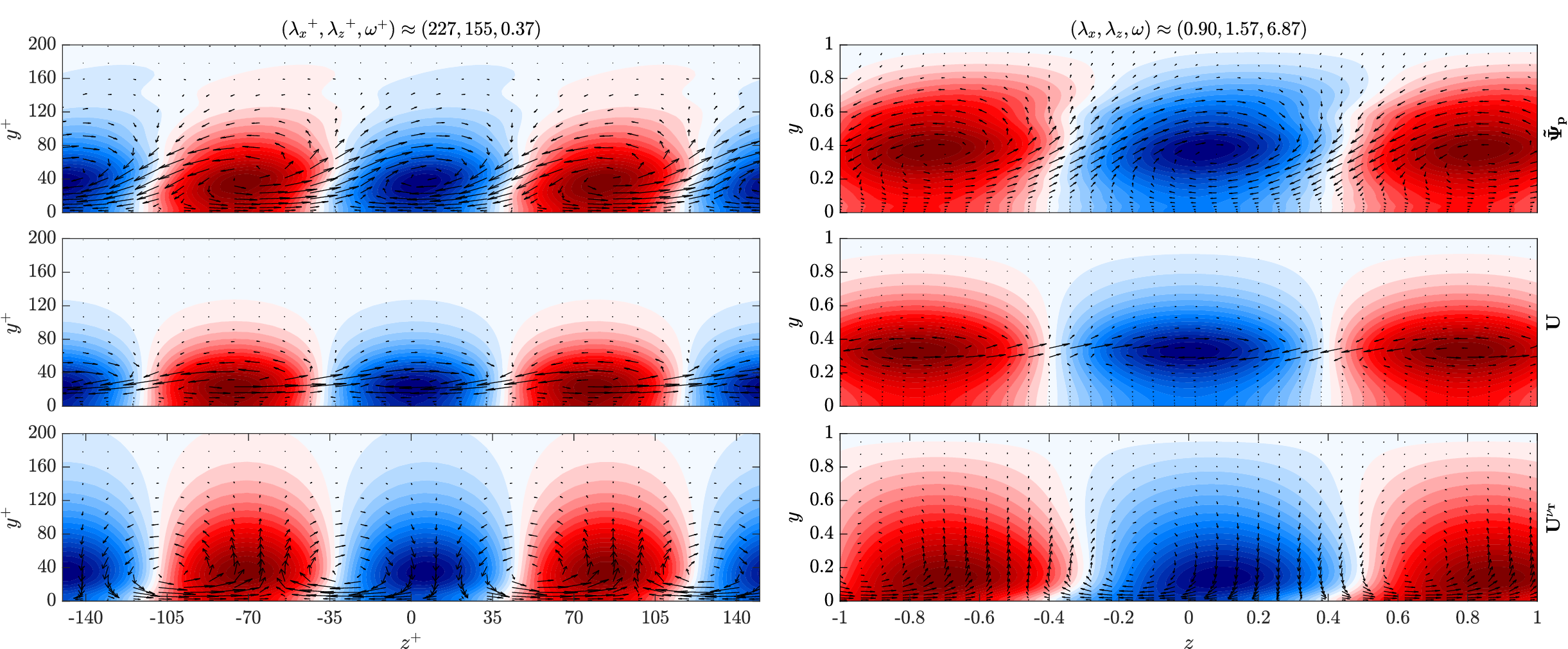}}
	\caption{Pressure near-wall (left frames) and large-scale (right frames) structures modes for the $Re_{\tau} \approx 550$ simulation. Rows, from top to bottom: first SPOD ($\boldsymbol{\tilde{\Psi}}_{p}$), resolvent ($\boldsymbol{U}$) and eddy viscosity resolvent ($\boldsymbol{U}^{{\nu_T}}$) modes. Contours correspond to the pressure component, with blue color associated to negative pressure and red color associated to positive pressure. Arrows denote wall-normal and spanwise velocity components.}
	\label{fig:p_spod_resolvent_modes_near-wall_large-scale_Retau550}
\end{figure}

SPOD ($\boldsymbol{\tilde{\Psi}}_{p}$), resolvent ($\boldsymbol{U}$) and eddy viscosity resolvent ($\boldsymbol{U}^{{\nu_T}}$) modes are shown in figure \ref{fig:p_spod_resolvent_modes_near-wall_large-scale_Retau550} for the near-wall and large-scale structures.
For brevity, only the first SPOD and the optimal resolvent and eddy viscosity resolvent modes are shown.
The pressure component is depicted in the contours, where blue color denotes negative values, whereas the red color indicates positive values in the color map scale.
For all frames, the arrows correspond to the wall-normal and spanwise velocity components; although SPOD targets pressure fluctuations here, the velocity components may be obtained by their inclusion in the realization matrix $\boldsymbol{\hat{G}}$ in equation \ref{eq:spod_true}, in a procedure corresponding to extended POD \citep{boree2003extended}.
Near-wall pressure structures (left frames of figure \ref{fig:p_spod_resolvent_modes_near-wall_large-scale_Retau550}) and the associated quasi-streamwise vortices peak at $y^+ \approx 35$, 20 and 40 for the $\boldsymbol{\tilde{\Psi}}_{p}$, $\boldsymbol{U}$ and $\boldsymbol{U}^{{\nu_T}}$ modes, respectively.
The quasi-streamwise vortices for the SPOD ($\boldsymbol{\tilde{\Psi}}_{p}$) and resolvent ($\boldsymbol{U}$) modes have clockwise rotation associated with negative (blue) pressure, whereas counter-clockwise vortices are associated with positive (red) pressure values.
Although the position of quasi-streamwise vortices (i.e. the phase with respect to pressure fluctuations) is correct for the resolvent mode, the vortices are more concentrated around the critical layer, as usual for resolvent modes with molecular viscosity \cite{morra2019relevance}.
This mode is similar to that shown by \citet{abreu2020resolvent}, who employed an L$_2$ norm accounting for the three velocity components (kinetic energy norm) to extract the SPOD and resolvent modes at $({\lambda_{x}}^{+},~{\lambda_{z}}^{+},~{\lambda_{t}}^{+}) \approx (1000,~100,~100)$, i.e. the wavenumber-frequency triplet of the premultiplied spectrum peak for the near-wall structures.
The large-scale structures span almost the entire channel half-height and have a dynamics similar to that observed for the near-wall structures.
For the SPOD and resolvent modes, the quasi-streamwise vortices have support at $y \approx 0.40$ and 0.35, respectively.
The eddy viscosity resolvent mode, on the other hand, peaks at $y \approx 0.15$ and also shows the quasi-streamwise vortices in the intersection between regions of high and low pressure.

\begin{figure}[!h]
	\centerline{\includegraphics[width=\textwidth]{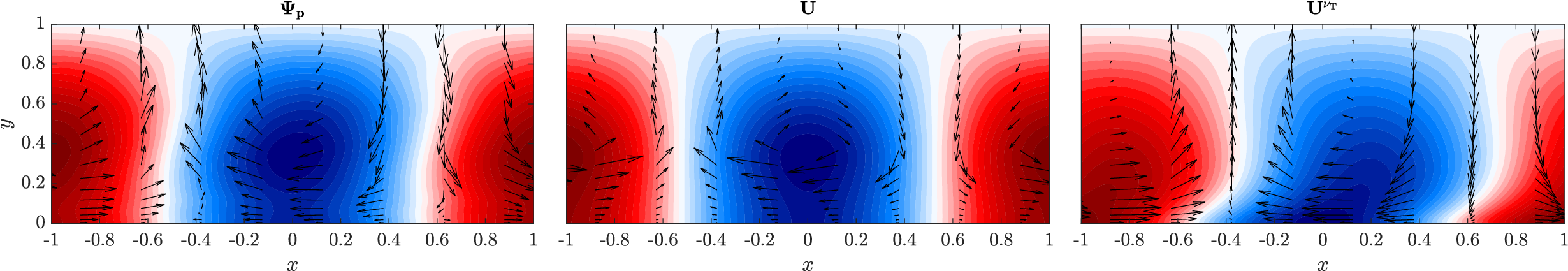}}
	\caption{Pressure spanwise-coherent structures for the $Re_{\tau} \approx 550$ simulation. Frames, from left to right: first SPOD ($\boldsymbol{\tilde{\Psi}}_{p}$), resolvent ($\boldsymbol{U}$) and eddy viscosity resolvent ($\boldsymbol{U}^{{\nu_T}}$) modes. See the comments in the caption of figure \ref{fig:p_spod_resolvent_modes_near-wall_large-scale_Retau550}.}
	\label{fig:p_spod_resolvent_modes_spanwise-coherent_Retau550}
\end{figure}

The pressure modes for the spanwise-coherent structures are shown in figure \ref{fig:p_spod_resolvent_modes_spanwise-coherent_Retau550}.
First, second and third columns correspond to the SPOD ($\boldsymbol{\tilde{\Psi}}_{p}$) resolvent ($\boldsymbol{U}$) and eddy viscosity resolvent ($\boldsymbol{U}^{{\nu_T}}$) modes, respectively.
The pressure structures span the channel half-height.
Moreover, for the three modal decompositions, regions of low pressure are associated with clockwise vortices, whereas for the regions of high pressure, the vortices rotate in the counter-clockwise direction.
Near the wall, the pressure component for the $\boldsymbol{U}^{{\nu_T}}$ modes displays a slight inclination towards the right direction, which is not observed for the $\boldsymbol{\tilde{\Psi}}_{p}$ and $\boldsymbol{U}$ modes.
Similarly to near-wall and large-scale structures, both resolvent modes display a fairly close agreement with the SPOD mode taken from the DNS.

\begin{figure}[!h]
	\centerline{\includegraphics[width=\textwidth]{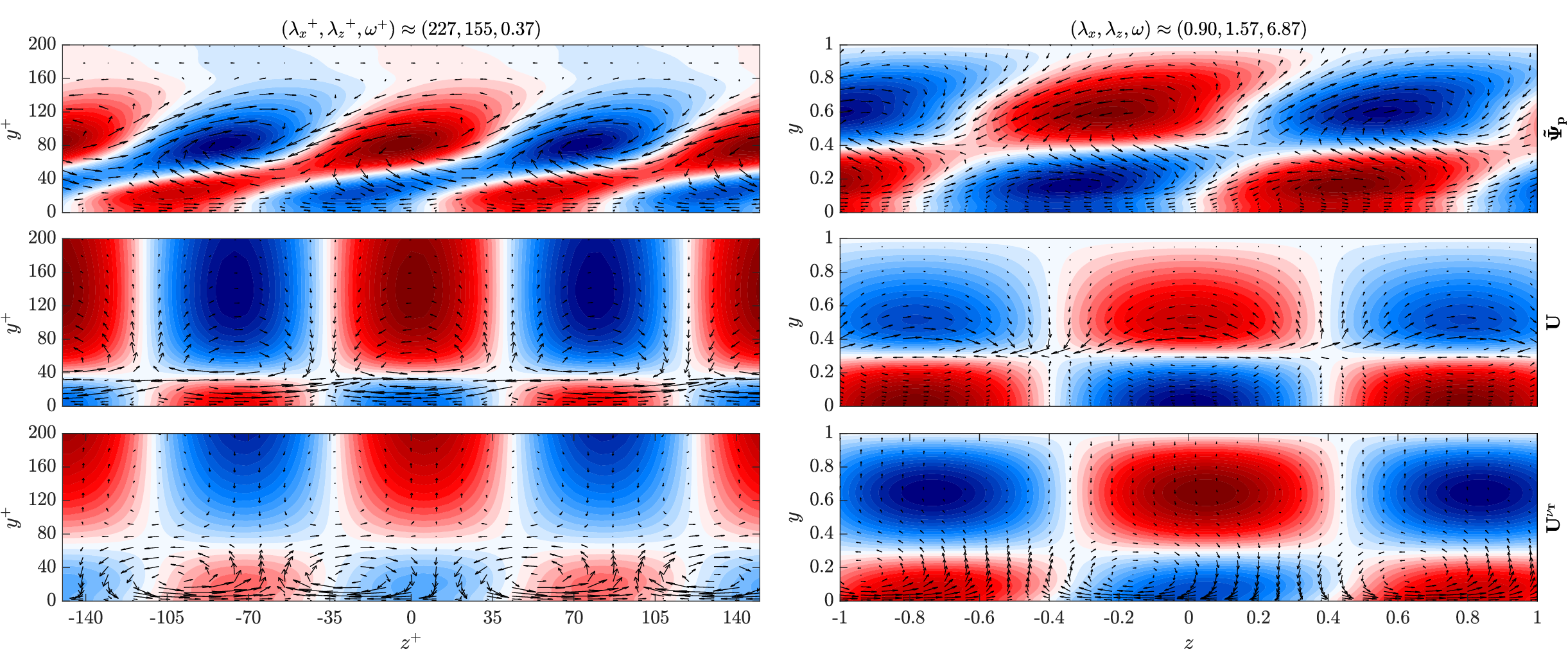}}
	\caption{Pressure near-wall (left frames) and large-scale (right frames) structures modes for the $Re_{\tau} \approx 550$ simulation. Rows, from top to bottom: second SPOD ($\boldsymbol{\tilde{\Psi}}_{p}$), resolvent ($\boldsymbol{U}$) and eddy viscosity resolvent ($\boldsymbol{U}^{{\nu_T}}$) modes. See the comments in the caption of figure \ref{fig:p_spod_resolvent_modes_near-wall_large-scale_Retau550}}.
	\label{fig:p_spod_resolvent_modes2_near-wall_large-scale_Retau550}
\end{figure}

\begin{figure}[!h]
	\centerline{\includegraphics[width=\textwidth]{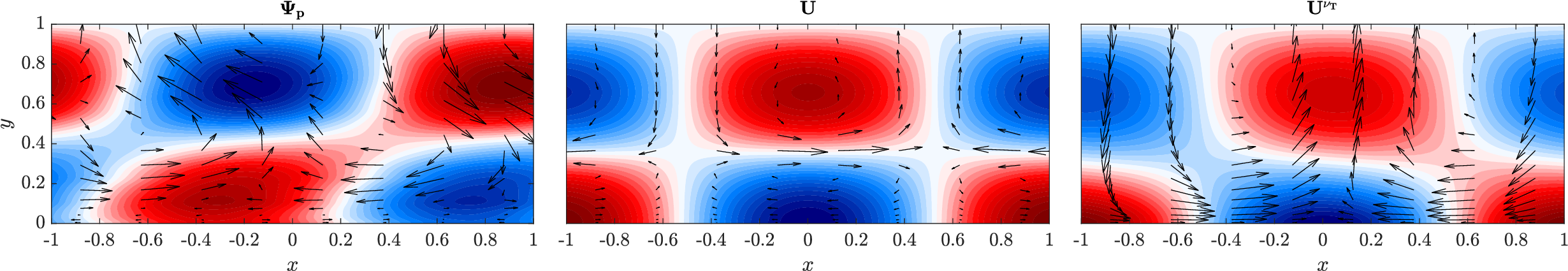}}
	\caption{Pressure spanwise-coherent structures for the $Re_{\tau} \approx 550$ simulation. Frames, from left to right: second SPOD ($\boldsymbol{\tilde{\Psi}}_{p}$), resolvent ($\boldsymbol{U}$) and eddy viscosity resolvent ($\boldsymbol{U}^{{\nu_T}}$) modes. See the comments in the caption of figure \ref{fig:p_spod_resolvent_modes_near-wall_large-scale_Retau550}.}
	\label{fig:p_spod_resolvent_modes2_spanwise-coherent_Retau550}
\end{figure}

Figures \ref{fig:p_spod_resolvent_modes2_near-wall_large-scale_Retau550} and \ref{fig:p_spod_resolvent_modes2_spanwise-coherent_Retau550} show the second resolvent and SPOD modes, in the same fashion as the figures 12 and 13.
We observe that the second mode now shows a pair of structures along $y$, in phase opposition.
This is similar to what is observed for corresponding velocity modes \citep{abreu2020resolvent}.
The resolvent modes reproduce the phase opposition observed in the SPOD mode, but with a sharper jump that contrasts with the smoother phase change along y in the SPOD mode.
The second modes appear to be related to a similar mechanism in the leading flow response, this time leading to structures in phase opposition along $y$.

\subsection{Comparison to velocity modes}
\label{sec:comparison_velocity}

\begin{figure}[!h]
	\centerline{\includegraphics[width=\textwidth]{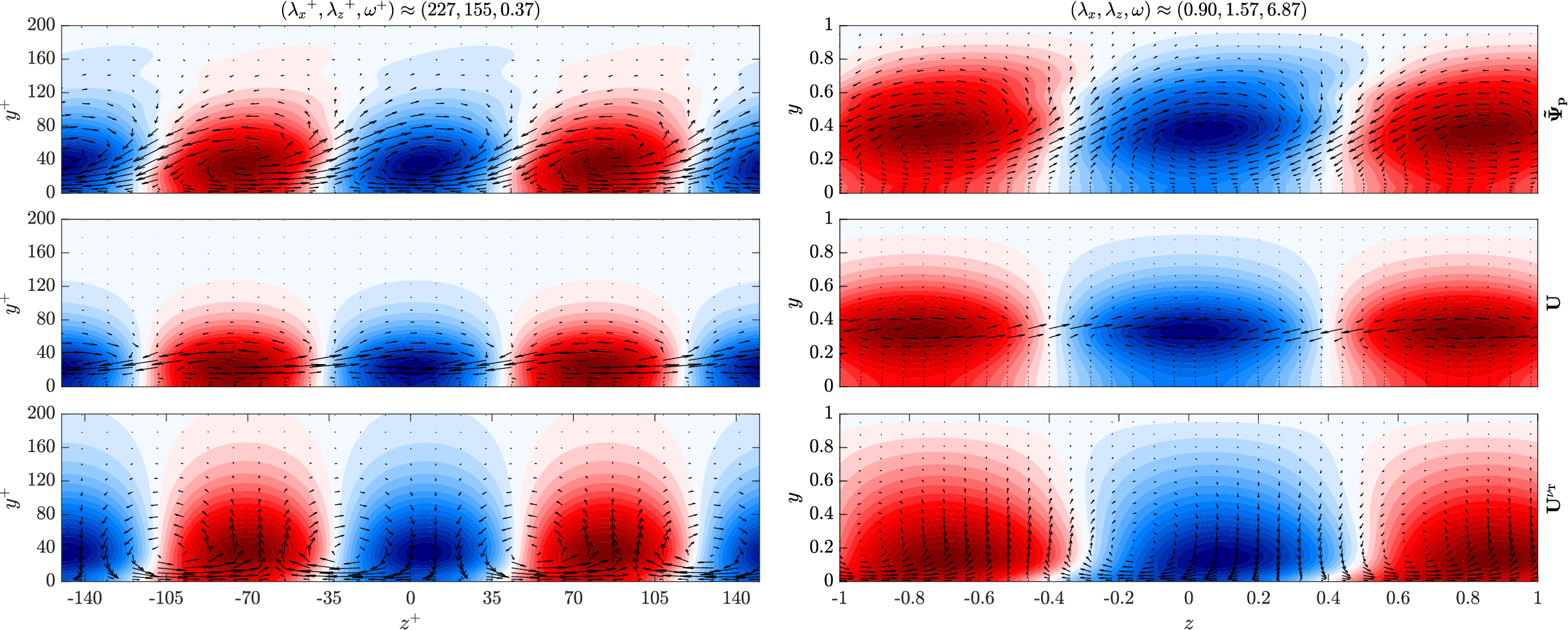}}
	\caption{TKE near-wall (left frames) and large-scale (right frames) structures modes for the $Re_{\tau} \approx 550$ simulation. See the comments in the caption of figure \ref{fig:p_spod_resolvent_modes_near-wall_large-scale_Retau550}.}
	\label{fig:TKE_spod_resolvent_modes_near-wall_large-scale_Retau550}
\end{figure}

\begin{figure}[!h]
	\centerline{\includegraphics[width=\textwidth]{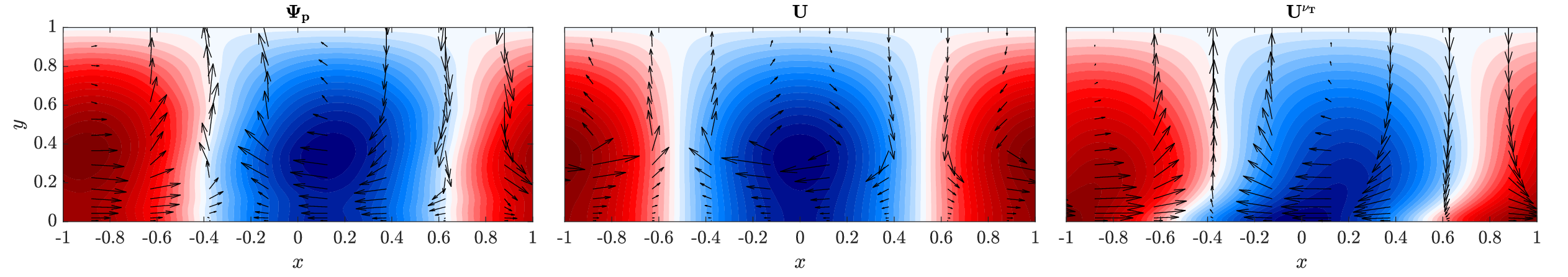}}
	\caption{TKE spanwise-coherent structures for the $Re_{\tau} \approx 550$ simulation. See the comments in the caption of figure \ref{fig:p_spod_resolvent_modes_spanwise-coherent_Retau550}.}
	\label{fig:TKE_spod_resolvent_modes_spanwise-coherent_Retau550}
\end{figure}

We now compare some features of the observed pressure structures with modes obtained considering the more usual metrics related to velocity fluctuations.
Figures \ref{fig:TKE_spod_resolvent_modes_near-wall_large-scale_Retau550} and \ref{fig:TKE_spod_resolvent_modes_spanwise-coherent_Retau550} display the SPOD ($\boldsymbol{\tilde{\Psi}}_{TKE}$), resolvent ($\boldsymbol{U}_{TKE}$) and eddy viscosity resolvent (${\boldsymbol{U}^{{\nu_T}}}_{TKE}$) modes for the near-wall and large scale structures and spanwise-coherent structure, respectively, defining operators $\boldsymbol{C}$ (equation \ref{eq:observation_matrix}) and $\boldsymbol{W}$ (equation \ref{eq:spod_pressure_weights}) to target the turbulent kinetic energy (TKE) as
\begin{equation}
	\boldsymbol{C} =
	\left[\begin{array}{cccc}
		\boldsymbol{I}	& \boldsymbol{Z}	& \boldsymbol{Z}	& \boldsymbol{Z}	\\
		\boldsymbol{Z}	& \boldsymbol{I}	& \boldsymbol{Z}	& \boldsymbol{Z}	\\
		\boldsymbol{Z}	& \boldsymbol{Z}	& \boldsymbol{I}	& \boldsymbol{Z}	\\
		\boldsymbol{Z}	& \boldsymbol{Z}	& \boldsymbol{Z}	& \boldsymbol{Z}	\\
	\end{array}\right] \mbox{,}
	\label{eq:observation_matrix_TKE}
\end{equation}
\noindent and
\begin{equation}
	\boldsymbol{W} =
	\left[\begin{array}{cccc}
		\boldsymbol{K}_{p}	& \boldsymbol{Z}			& \boldsymbol{Z}			& \boldsymbol{Z}	\\
		\boldsymbol{Z}			& \boldsymbol{K}_{p}	& \boldsymbol{Z}			& \boldsymbol{Z}	\\
		\boldsymbol{Z}			& \boldsymbol{Z}			& \boldsymbol{K}_{p}	& \boldsymbol{Z}	\\
		\boldsymbol{Z}			& \boldsymbol{Z}			& \boldsymbol{Z}			& \boldsymbol{Z}	\\
	\end{array}\right] \mbox{.}
	\label{eq:spod_pressure_weights_TKE}
\end{equation}
The obtained TKE-targeted structures display overall similarities the pressure-targeted structures (figures \ref{fig:p_spod_resolvent_modes_near-wall_large-scale_Retau550} and \ref{fig:p_spod_resolvent_modes_spanwise-coherent_Retau550}).
The TKE-SPOD modes are more inclined towards the right direction with respect to their pressure counterpart.
While the TKE and pressure resolvent modes are very similar, likely due to the low-rank nature of the linearized system, the eddy viscosity resolvent has different modes, with pressure structures with strong variations near the wall, in contrast with the smoother SPOD modes.
Moreover, the spanwise-coherent structure using pressure and TKE metrics are almost identical, except for the eddy viscosity resolvent mode that has a near wall behavior in contrast with the SPOD mode.
We observe that the use of the pressure as a single variable determining the level of agreement simplifies the modeling task, as a single scalar SPOD mode must be matched by linearized models.

\begin{figure}[!h]
	\centerline{\includegraphics[width=\textwidth]{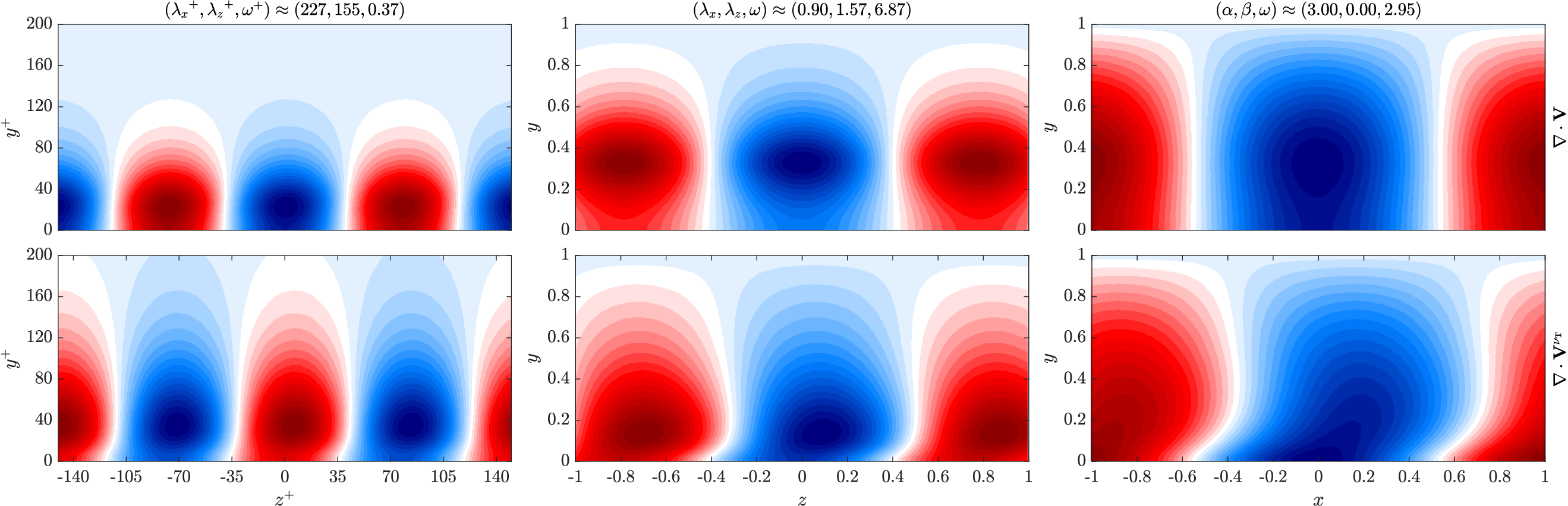}}
	\caption{Divergence of the first forcing mode (top row) and eddy-viscosity forcing mode (bottom row) for the $Re_{\tau} \approx 550$ case. Frames, from left to right: near-wall, large-scale and spanwise coherent structures.}
	\label{fig:divergence_resolvent_modes_spanwise-coherent_Retau550}
\end{figure}

To explore the different mechanisms leading to pressure and velocity responses, figure \ref{fig:divergence_resolvent_modes_spanwise-coherent_Retau550} shows the divergence of the resolvent forcing first mode evaluated with (bottom row) and without the eddy viscosity model (top row) for the three studied structures, i.e. near-wall, large-scale and spanwise coherent structures (columns, from left to right).
Differently from the forcing modes obtained with a TKE energy norm, which are solenoidal/incompressible \citep{morra2021colour}, the divergence of the forcing when accounting for a pressure energy norm is not zero.
Thus, there are additional mechanisms leading to pressure fluctuations, related to the non-zero divergence of non-linear terms.

\subsection{Parametric comparison between resolvent and SPOD modes}
\label{sec:comparison_resolvent_SPOD}

To assess a quantitative comparison between the SPOD and resolvent pressure modes, it is possible to project the first SPOD mode onto the optimal resolvent for a combination $(\alpha,~\beta,~\omega)$ in the same fashion as \citet{abreu2020spod}, i.e.
\begin{equation}
	\Gamma =  \left\lvert\frac{\langle \boldsymbol{a},~\boldsymbol{b} \rangle}{||\boldsymbol{a}||~||\boldsymbol{b}||}\right\rvert \mbox{,}
	\label{eq:projection}
\end{equation}
\noindent where $\Gamma$ is the normalized projection, with $\Gamma \in [0,~1]$, $\boldsymbol{a}$ and $\boldsymbol{b}$ are two modes to be compared, $\langle \cdot,\cdot \rangle$ denotes the L$_2$ inner product, $||\cdot||$ indicates the L$_2$ norm and $\left\lvert\cdot\right\rvert$ indicates absolute value.
It is important to remark that $\Gamma = 0$ and $\Gamma = 1$ indicate that the vectors are orthogonal and perfectly aligned, respectively.
Here, $\boldsymbol{a} = \boldsymbol{\tilde{\psi}}_1$, i.e. the first SPOD pressure mode, and $\boldsymbol{b} = \boldsymbol{U}_1$ or $\boldsymbol{b} = {{\boldsymbol{U}_1}^{{\nu_T}}}$, i.e. the optimal resolvent or eddy viscosity resolvent pressure mode.

\begin{figure}[!h]
	\centerline{\includegraphics[width=\textwidth]{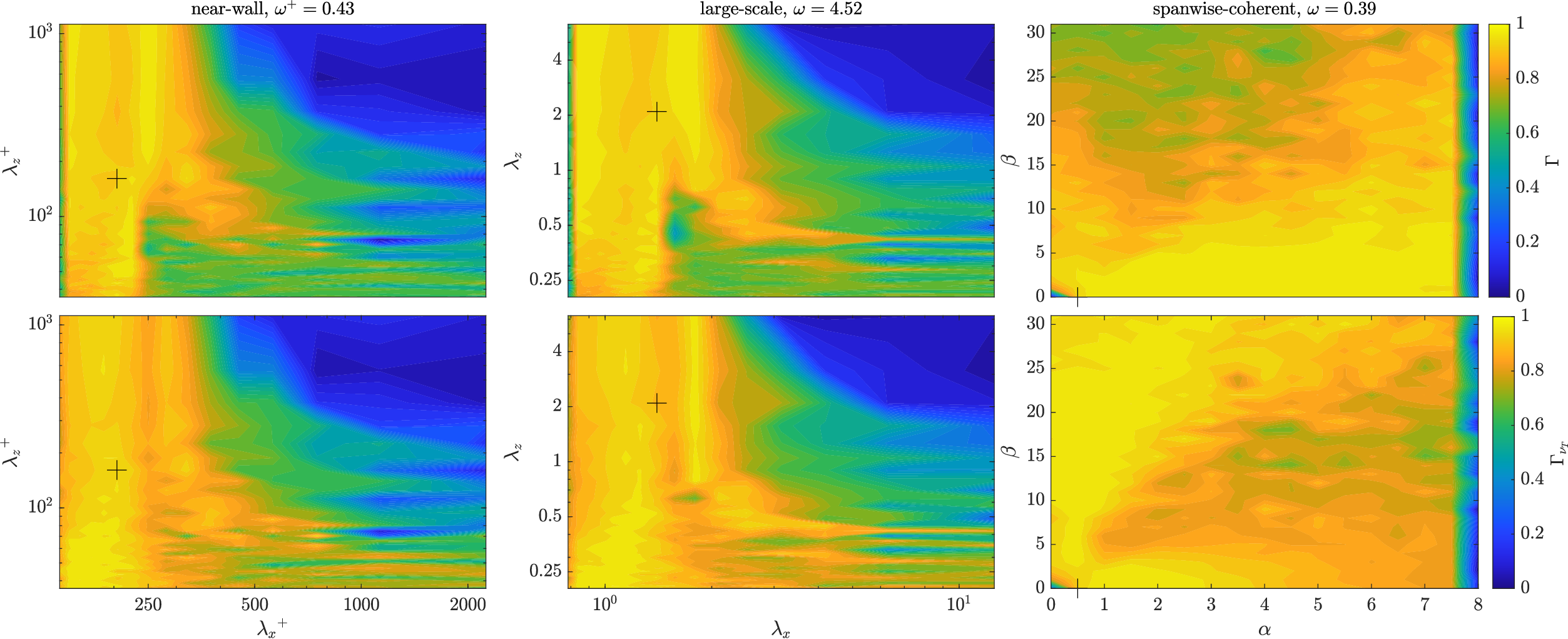}}
	\caption{Projection coefficients (pressure norm) of the first SPOD mode on the first resolvent (top row) and eddy viscosity resolvent (bottom row) modes for the $Re_{\tau} \approx 180$ case.}
	\label{fig:projection_Retau180}
\end{figure}

\begin{figure}[!h]
	\centerline{\includegraphics[width=\textwidth]{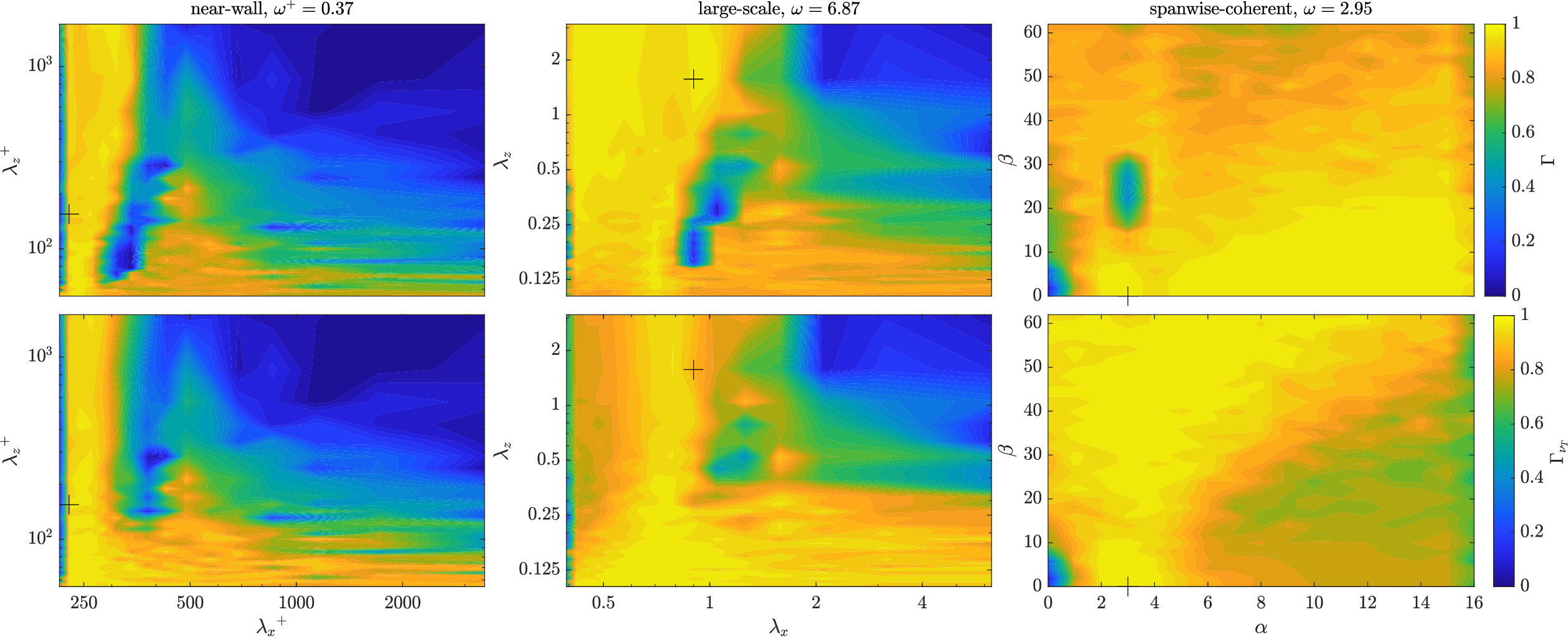}}
	\caption{Projection coefficients (pressure norm) of the first SPOD mode on the first resolvent (top row) and eddy viscosity resolvent (bottom row) modes for the $Re_{\tau} \approx 550$ case.}
	\label{fig:projection_Retau550}
\end{figure}

Figures \ref{fig:projection_Retau180} and \ref{fig:projection_Retau550} exhibit the projection coefficients for the $Re_{\tau} \approx 180$ and 550 cases, respectively, for the temporal frequencies of the near-wall, large-scale and spanwise-coherent structures.
The coefficients were evaluated for all wavenumbers contained in the database and are shown in contours in the figure; top row for the resolvent operator and bottom row for the eddy viscosity resolvent operator.
The cross markers on each frame indicate the wavenumber pair correspondent to the structure's energy peak.
Overall, for the frequency corresponding to near-wall structures, high projection coefficient levels are observed for low streamwise and spanwise wavelengths.
For the large-scale frequency, the band of high level coefficient levels is enlarged, but the higher levels are still attained for the lower wavelengths.
Regarding the characteristic frequency of spanwise coherent structures, the normalized projection levels are approximately homogeneous in the wavenumber space, despite a few regions where lower and higher levels can be observed.
However, for both Reynolds numbers, note the molecular viscosity resolvent leads to high projection coefficients in the high $\alpha$/low $\beta$ region, while the eddy viscosity formulation leads to higher coefficients in the low $\alpha$/high $\beta$ region.
Yet, for the $Re_{\tau} \approx 550$ case and considering molecular viscosity resolvent (top-right frame of figure \ref{fig:projection_Retau550}), there is a gap in the projection coefficient around  the $2 \leq \alpha \leq 4$ and $20 \leq \beta \leq 30$ regions, which does not occur for the eddy viscosity resolvent nor for the $Re_{\tau} \approx 180$ case and both resolvent formulations.
Overall, besides the considerations above, for the three considered frequencies, there is no clear advantage for resolvent models with or without the eddy viscosity, with both approaches having similar parameter ranges with good agreement with the leading SPOD modes.

Tables \ref{tab:projection_Retau180} and \ref{tab:projection_Retau550} display the projection coefficients for the $Re_{\tau} \approx 180$ and 550 cases, respectively, for the near-wall, large-scale and spanwise-coherent structures (cross markers in figure \ref{fig:projection_Retau550}).
It is observed that indeed the pressure SPOD first modes project very well onto the resolvent and eddy viscosity resolvent optimal pressure modes of the three selected structures, especially the near-wall and spanwise-coherent structures, with $\Gamma \geq 0.94$.
As commented before, there is no clear advantages for one of the resolvent models.

\begin{table}[!h]
  \begin{center}
		\def~{\hphantom{0}}
		\caption{Projection coefficients for the pressure modes and $Re_{\tau} \approx 180$ case.}
		\resizebox{\textwidth}{!}{
		\begin{tabular}{c c c c }
			\multirow{2}{*}{Projection}	& near-wall structure																										& large-scale structure  																			& spanwise-coherent structure								\\ [3pt]
																	& $({\lambda_x}^+,~{\lambda_z}^+,~{\omega}^+) \approx (204,~161,~0.43)$	& $(\lambda_x,~\lambda_z,~\omega) \approx (1.40,~2.09,~4.52)$	& $(\alpha,~\beta,~\omega) \approx (0.50,~0.00,~0.39)$	\\ [6pt]
			\hline \\ [3pt]
			$\Gamma$										& 0.94																																	& 0.91																												& 0.97														\\ [3pt]
			$\Gamma_{{\nu_T}}$						& 0.95																																	& 0.94																												& 0.96														\\ [3pt]
		\end{tabular}
		}
		\label{tab:projection_Retau180}
  \end{center}
\end{table}

\begin{table}[!h]
  \begin{center}
		\def~{\hphantom{0}}
		\caption{Projection coefficients for the pressure modes and $Re_{\tau} \approx 550$ case.}
		\resizebox{\textwidth}{!}{
		\begin{tabular}{c c c c }
			\multirow{2}{*}{Projection}	& near-wall structure																										& large-scale structure  																			& spanwise-coherent structure								\\ [3pt]
																	& $({\lambda_x}^+,~{\lambda_z}^+,~{\omega}^+) \approx (227,~155,~0.37)$	& $(\lambda_x,~\lambda_z,~\omega) \approx (0.90,~1.57,~6.87)$	& $(\alpha,~\beta,~\omega) \approx (3.00,~0.00,~2.95)$	\\ [6pt]
			\hline \\ [3pt]
			$\Gamma$										& 0.96																																	& 0.89																												& 0.99														\\ [3pt]
			$\Gamma_{{\nu_T}}$						& 0.98																																	& 0.97																												& 0.98														\\ [3pt]
		\end{tabular}
		}
		\label{tab:projection_Retau550}
  \end{center}
\end{table}

\begin{figure}[!h]
	\centerline{\includegraphics[width=\textwidth]{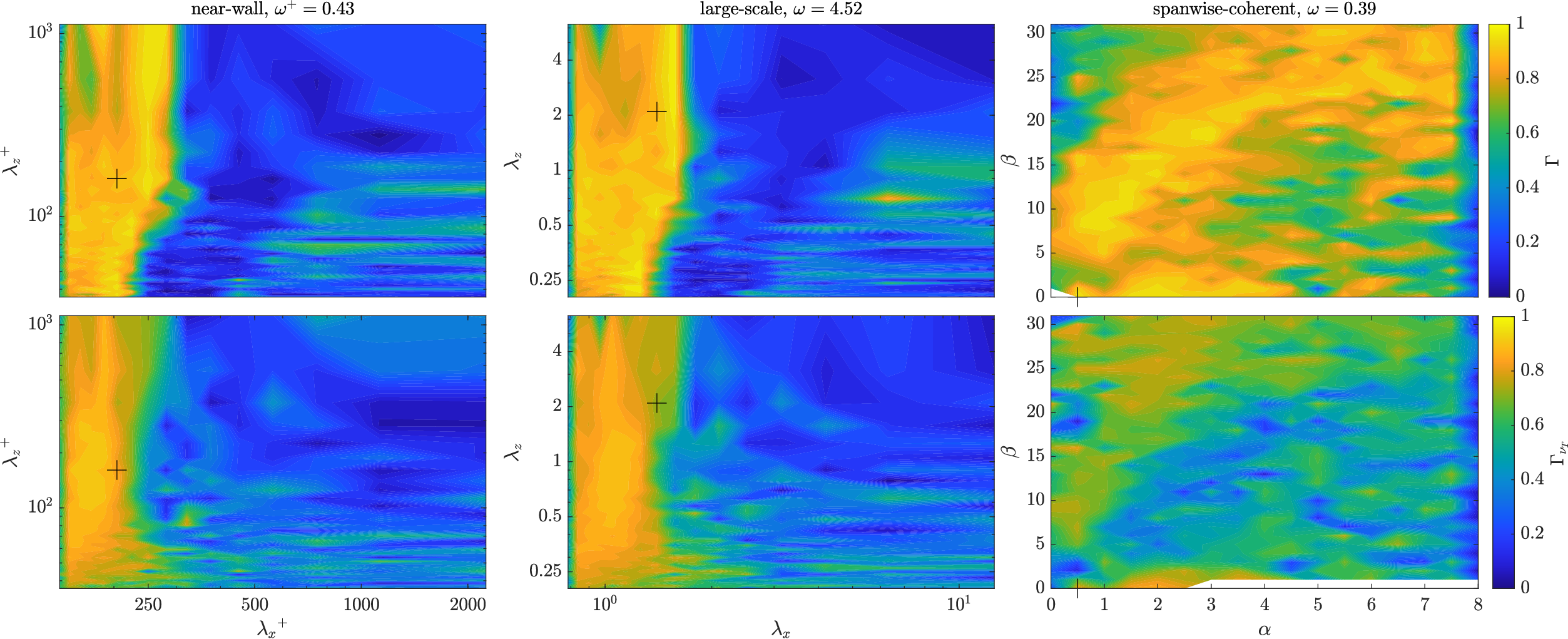}}
	\caption{Projection coefficients (TKE norm) of the first SPOD mode on the first resolvent (top row) and eddy viscosity resolvent (bottom row) modes for the $Re_{\tau} \approx 180$ case.}
	\label{fig:projectionTKE_Retau180}
\end{figure}

\begin{figure}[!h]
	\centerline{\includegraphics[width=\textwidth]{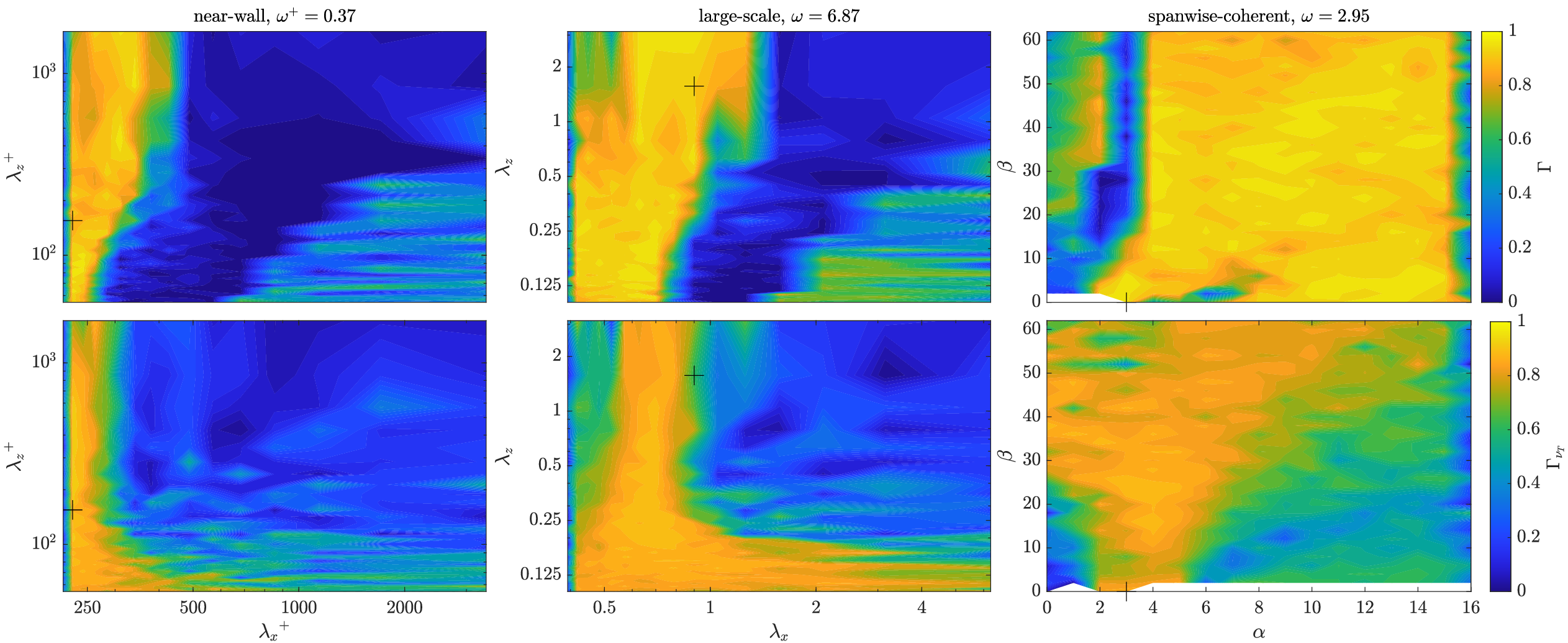}}
	\caption{Projection coefficients (TKE norm) of the first SPOD mode on the first resolvent (top row) and eddy viscosity resolvent (bottom row) modes for the $Re_{\tau} \approx 550$ case.}
	\label{fig:projectionTKE_Retau550}
\end{figure}

Figures \ref{fig:projectionTKE_Retau180} and \ref{fig:projectionTKE_Retau550} show the projection coefficients using the TKE norm instead of the pressure norm, as in \citet{abreu2020resolvent, abreu2020spod}.
In comparison to the pressure norm counterpart, figures \ref{fig:projection_Retau180} and \ref{fig:projection_Retau550}, the region of high projection coefficient values is smaller for the three studied structures.
Regarding the near-wall and spanwise-coherent structures, higher values are obtained only for the smaller streamwise wavelengths, whereas for the spanwise-coherent structure, the projection coefficient is more uniformly spread along the streamwise and spanwise wavelength space, as for the pressure norm.
The differences between the resolvent and the eddy viscosity resolvent when using the TKE norm are qualitatively the same as when using the pressure norm, i.e. overall lower projection levels are attained for the eddy viscosity resolvent, although the region comprising low spanwise wavenumbers shows higher projection coefficient levels for higher streamwise wavenumbers.

\section{Role of different forcing components}
\label{sec:input_output_analysis}

A further effort to explore how the forcing leads to the observed pressure statistics is pursued in this section.
In order to do so, let us consider the effect of the input sub-blocks (forcing CSD) on the output (pressure CSD).
The CSD $\boldsymbol{F}$ can be interpreted as a sum of sub-blocks ($\boldsymbol{F}_{ij}$), with each sub-block accounting for a forcing component, i.e.
\begin{equation}
	\boldsymbol{F} = \boldsymbol{F}_{xx} + \boldsymbol{F}_{xy} + \boldsymbol{F}_{xz} + \boldsymbol{F}_{yy} + \boldsymbol{F}_{yx} + \boldsymbol{F}_{yz} + \boldsymbol{F}_{zz} + \boldsymbol{F}_{zx} + \boldsymbol{F}_{zy} \mbox{,}
	\label{eq:forcing_subblocks}
\end{equation}
\noindent where the subscripts $i$ and $j$ in the $\boldsymbol{F}_{ij}$ sub-block notation indicate the streamwise ($x$), wall-normal ($y$) and/or spanwise ($z$) forcing components. 
Note that because input $\boldsymbol{F}$ is Hermitian for $i \ne j$, we will consider $\boldsymbol{F}_{ij}$ and $\boldsymbol{F}_{ji}$ combined.

\begin{figure}[!h]
	\centerline{\includegraphics[width=\textwidth]{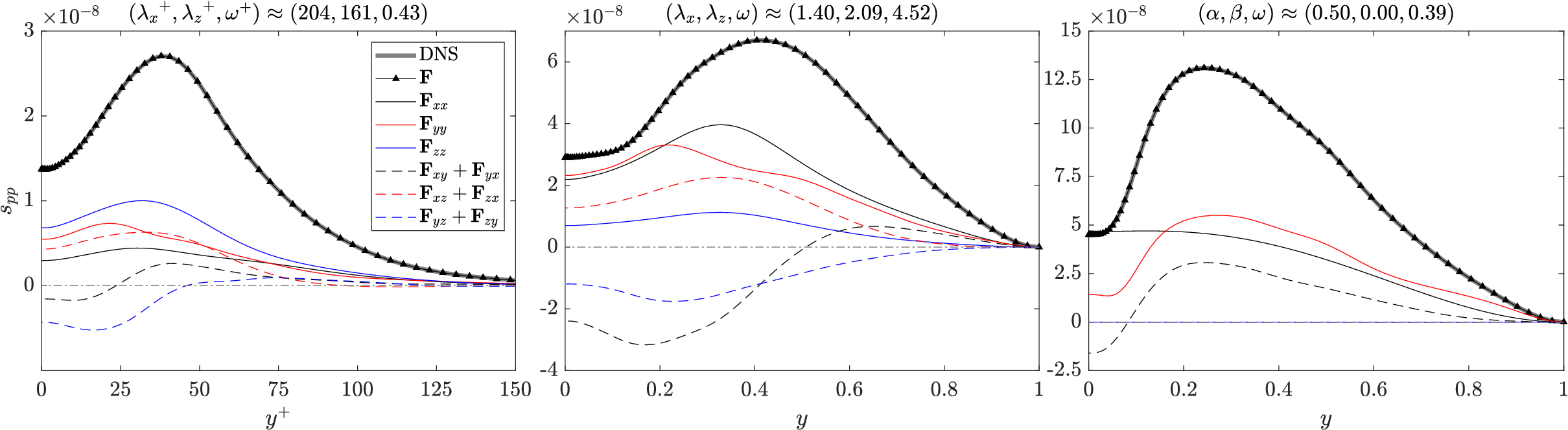}}
	\caption{Pressure component response $\boldsymbol{S} = \boldsymbol{H} \boldsymbol{F}_{ij} \boldsymbol{H}^{\dagger}$ from sub-blocks $\boldsymbol{F}_{ij}$ for $Re_{\tau} \approx 180$. Solid lines: response from sub-blocks $\boldsymbol{F}_{ii}$. Dashed lines: response from sub-blocks $\boldsymbol{F}_{ij}$ with $i \neq j$. Triangles: response from full forcing CSD. Gray thick line: response from DNS. Frames, from left to right: near-wall, large-scale and spanwise-coherent structures, respectively.}
	\label{fig:Pff_subblocks_p_y0_y05_Retau180}
\end{figure}

\begin{figure}[!h]
	\centerline{\includegraphics[width=\textwidth]{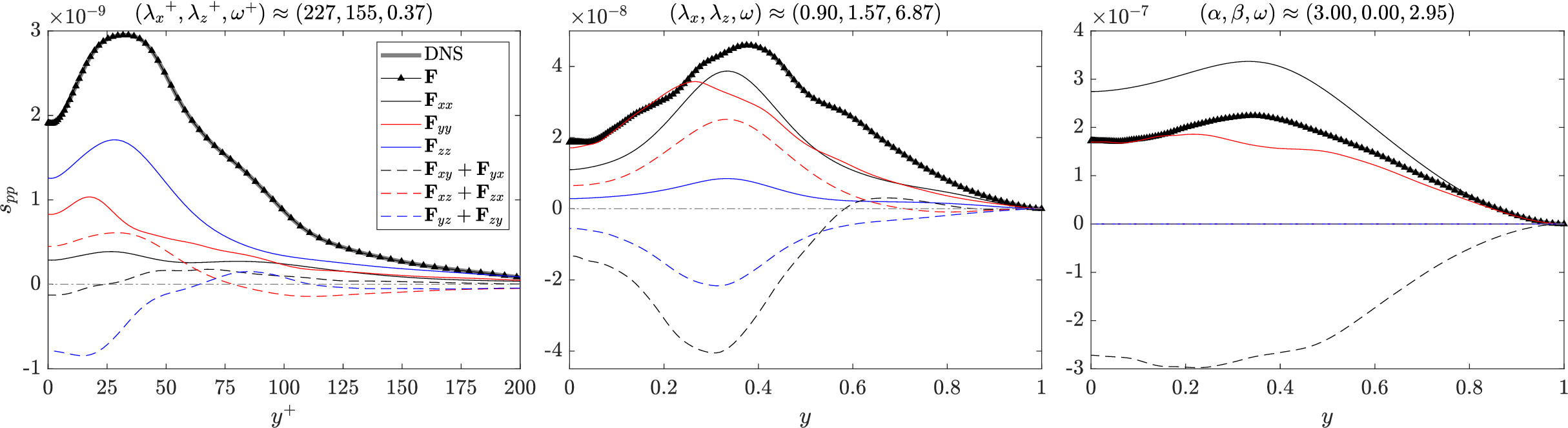}}
	\caption{Pressure component response $\boldsymbol{S} = \boldsymbol{H} \boldsymbol{F}_{ij} \boldsymbol{H}^{\dagger}$ from sub-blocks $\boldsymbol{F}_{ij}$ for $Re_{\tau} \approx 550$. See the comments in the caption of figure \ref{fig:Pff_subblocks_p_y0_y05_Retau180}.}
	\label{fig:Pff_subblocks_p_y0_y05_Retau550}
\end{figure}

Figures \ref{fig:Pff_subblocks_p_y0_y05_Retau180} and \ref{fig:Pff_subblocks_p_y0_y05_Retau550} show the output $\boldsymbol{s}_{pp}$ as a function of the inputs $\boldsymbol{F}_{ij}$ and $\boldsymbol{F}$ for $Re_{\tau} \approx 180$ and 550 simulations, respectively.
The left frames of the figures correspond to the near-wall structures, whereas the middle and right frames denote large-scale and spanwise-coherent structures, respectively.
A first observation is that using the full forcing statistics $\boldsymbol{F}$ leads to a nearly exact recovery of the full response statistics $\boldsymbol{S}$ from the DNS.
This provides further validation of the present database.

The forcing CSD terms considering $i = j$ provide positive contribution of pressure PSD.
For the near-wall (of both cases) and large-scale (of the $Re_{\tau} \approx 180$ case) structures, the $\boldsymbol{F}_{xy}$ and $\boldsymbol{F}_{yz}$ terms have a slightly negative contribution to the pressure output, in a region very close to the wall, whereas the $\boldsymbol{F}_{xz}$ terms provide a slightly positive contribution.
For the large-scale structure for the $Re_{\tau} \approx 550$ channel, the $\boldsymbol{F}_{xy}$ and $\boldsymbol{F}_{yz}$ terms have a negative contribution at $y \approx 0.3$, whereas the $\boldsymbol{F}_{xz}$ term has a positive contribution around this same wall-normal position.
Regarding the spanwise-coherent structures, all terms related to the spanwise forcing presented no contribution to the output, as expected; the pressure structures are thus two-dimensional, resulting solely from streamwise and wall-normal forcing.

The scenario presented here contrasts with that of \citet{morra2021colour}, who studied the contributions of the input CSD sub-blocks on the streamwise velocity component CSD output for near-wall and large-scale structures.
\citet{morra2021colour} showed that when $i = j$, the CSD terms positively contribute to the streamwise velocity output, whereas the $i \neq j$ terms mostly presented a negative contribution.
The authors observed that both positive and negative contributions have approximately the same amplitude levels, which lead to cancellations in the output if the full input (forcing) CSD is considered.
On the other hand, overall, in the present study the positive contributions of the input CSD sub-blocks have higher amplitudes than the negative contributions.
Therefore, when all input CSD terms are considered, a positive pressure output is obtained, with higher amplitude than those observed if only isolated input CSD sub-blocks are considered, in a mostly constructive interference.
Differently from \citet{morra2021colour}, it seems that for the pressure component the input CSD off-diagonal terms ($i \neq j$) do not play a leading role, at least for the near-wall structures, as the magnitude of such terms is much smaller than that of the diagonal terms ($i = j$).
The exceptions seem to occur for the large-scale and spanwise-coherent structures at $Re_{\tau} \approx 550$ (middle and right frames of figure \ref{fig:Pff_subblocks_p_y0_y05_Retau550}), for which the $\boldsymbol{F}_{xy}$ term has a large negative amplitude that significantly contributes for the output when considering the full forcing CSD.
Such characteristics of non-linear terms suggest that a simpler forcing model may be possible if one is interested in pressure structures.
The lower level of destructive interference also helps understanding the good performance of both resolvent modes in predicting the leading SPOD modes.

\section{Conclusions}
\label{sec:conclusions}

Pressure fluctuations in turbulent channel flows at $Re_{\tau} \approx 180$ and 550 were studied in this paper, using results from direct numerical simulations (DNSs).
The simulation were validated against literature results and good agreement was achieved for the rms \citep{delalamo2003spectra} and one-dimensional power spectra in the streamwise and spanwise directions \citep{choi1990space, anantharamu2020analysis}.
The two databases allowed extraction of dominant pressure structures, analyzed using wavenumber spectra and spectral proper orthogonal decomposition (SPOD).
SPOD modes were then compared to results from resolvent analysis in order to evaluate the potential of linearized models to represent such pressure structures.

Peak wavenumber-frequency combinations were selected based on spectra without and with pre-multiplication for planes at the channel wall ($y = 0$) and at $y \approx 0.5$.
Three characteristic structures were studied in more detail, corresponding to near-wall, large-scale and spanwise-coherent ($\beta = 0$) structures.
The premultiplied spectra for the wall plane (near-wall structures) showed a peak that collapsed in inner units at $({\lambda_x}^+,~{\lambda_z}^+,~\omega^+) \approx (200,~160,~0.4)$ for both Reynolds numbers.
Regarding the large-scale structures ($y \approx 0.5$), the premultiplied spectra peak do not collapse for the two Reynolds numbers.

Comparisons among the pressure SPOD and resolvent modes, with and without the inclusion of an eddy viscosity model, were performed.
Both resolvent and eddy viscosity resolvent modes displayed good agreement with the SPOD modes.
Such results can be explained by a clearer low-rank nature of the pressure SPOD modes, with a strong dominance of the leading mode for the wavenumber-frequency combinations presently studied.
As the resolvent modes comprise quasi-streamwise (for near-wall and large-scale structures) and spanwise vortices with pressure peaking at vortex centers, the analysis shows that forcing such vortices is the dominant mechanism leading to pressure fluctuations in turbulent channel flows.
Differently from works that focused on velocity components \citep{morra2021colour, symon2023eddy}, the eddy viscosity resolvent pressure modes does not show a clear advantage in terms of better agreement with the SPOD pressure modes.

The analysis of the nonlinear terms and their individual impacts on the pressure output reveals that the forcing terms have a predominantly constructive contribution to the output.
Moreover, with exception of the large-scale structures of the $Re_{\tau} \approx 550$ channel, the off-diagonal terms of the forcing CSD (see expression \ref{eq:forcing_subblocks}), are less important than the diagonal terms to obtain a good prediction of the pressure outputs.

In a previous study by our group, which employed a resolvent-based formulation to estimate the flow field from wall measurements \citep{amaral2021resolvent}, we noted that pressure structures were more accurate predicted than the velocity components, especially for distances far from the wall, with less dependence on the details of the forcing considered in the linearized model.
The present results show a more marked low-rank behavior of pressure fluctuations, which may explain the ability of resolvent models to predict or estimate pressure structures.
We believe that the results shown in this paper reinforces the need for proper modeling the color of forcing (nonlinear) terms when the problem is cast in the input-output framework, albeit with a lower effect of forcing color on pressure fluctuations compared to velocity components.
Moreover, the resolvent operator is a valuable asset to study the most energetic pressure structures and obtain relevant amplification mechanisms for turbulent pressure fluctuations.


\noindent{\bf Acknowledgements\bf{.}} \\ The authors are indebted to anonymous reviewers that provided an excellent feedback to enhance the quality of this study. \\

\noindent{\bf Funding\bf{.}} F. R. do Amaral received funding from from São Paulo Research Foundation (FAPESP/Brazil), grant \#2019/02203-2. A. V. G. Cavalieri was supported by the National Council for Scientific and Technological Development (CNPq/Brazil), grant \#313225/2020-6. The authors were also funded by FAPESP/Brazil, grant \#2019/27655-3. \\

\noindent{\bf  Author ORCID\bf{.}} F. R. do Amaral, https://orcid.org/0000-0003-1158-3216; A. V. G. Cavalieri, https://orcid.org/0000-0003-4283-0232. \\

\noindent{\bf Declaration of Interests\bf{.}} The authors report no conflict of interest. \\


\appendix

\section{Complementary results for the $Re_{\tau} \approx 180$ simulation}
\label{app:Retau180_results}

\begin{figure}[!h]
	\centerline{\includegraphics[width=\textwidth]{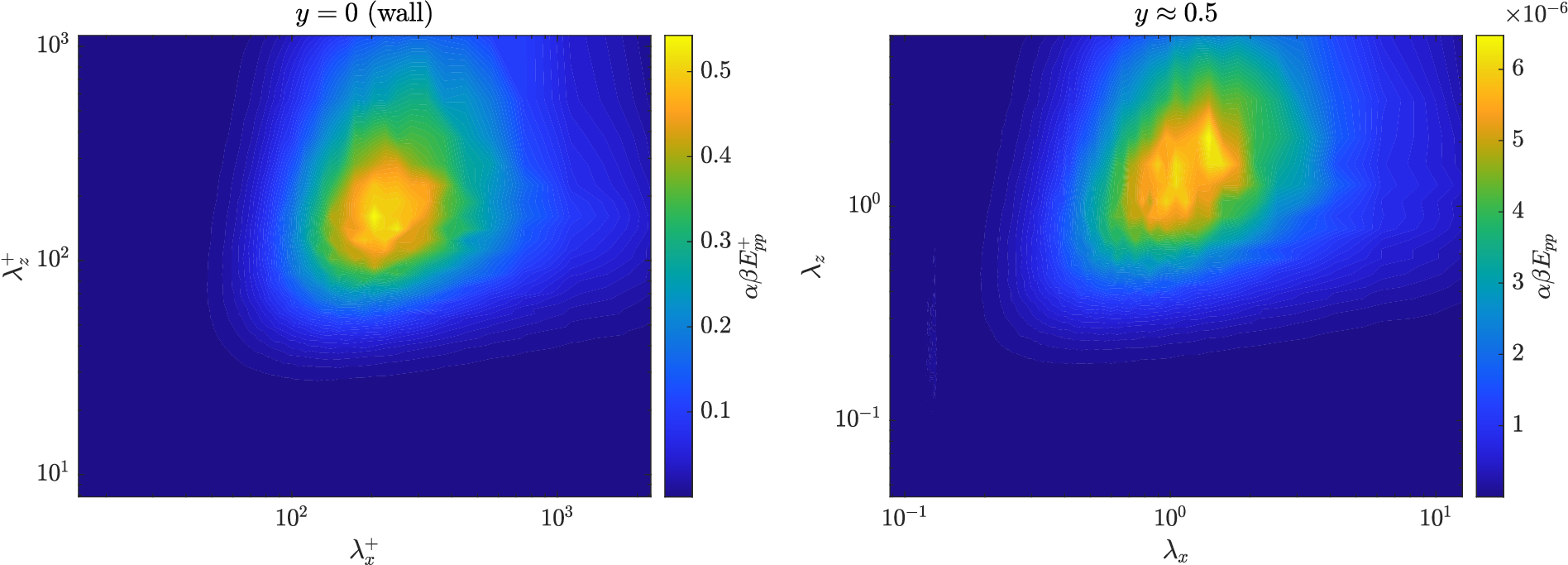}}
	\caption{Premultiplied energy spectra for $Re_{\tau} \approx 180$ simulation.}
	\label{fig:pmspectra_yp15_y05_Retau180}
\end{figure}

\begin{figure}[!h]
	\centerline{\includegraphics[width=\textwidth]{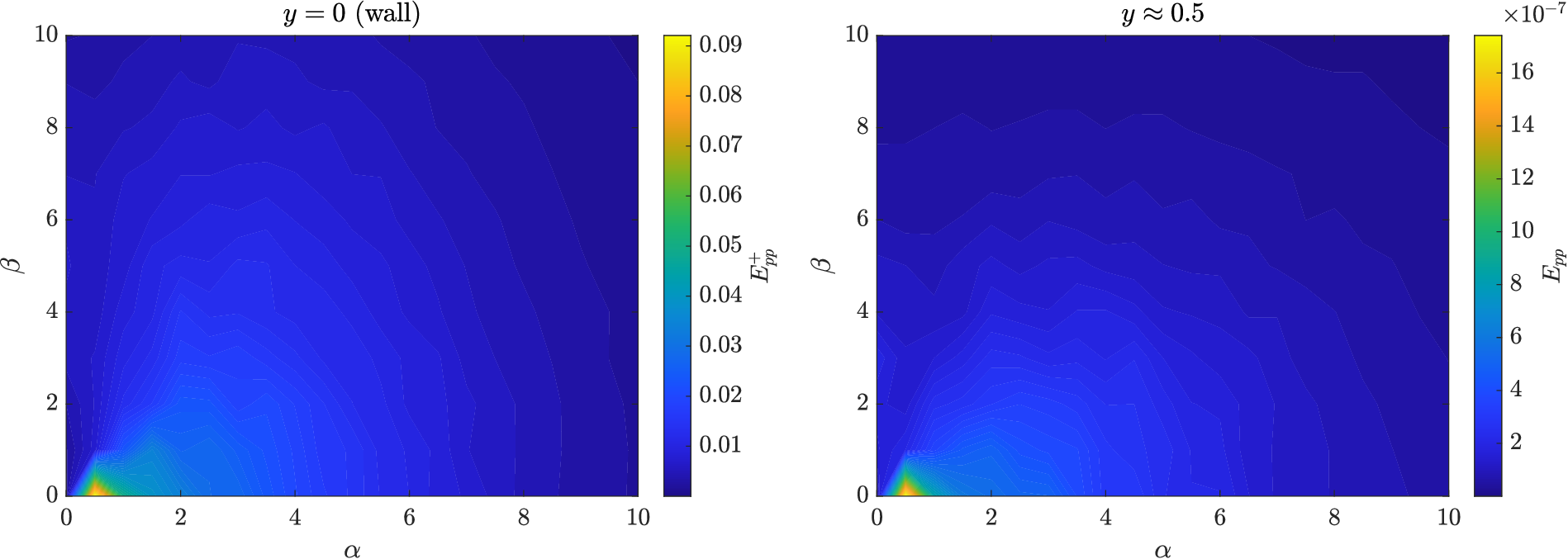}}
	\caption{Energy spectra for $Re_{\tau} \approx 180$ simulation.}
	\label{fig:spectra_y0_y05_Retau180}
\end{figure}

Figure \ref{fig:pmspectra_yp15_y05_Retau180} show the spectra evaluated with premultiplication.
The ${\lambda_x}^+$ peak value agrees with previous studies \citep{panton2017correlation, anantharamu2020analysis}.
At a distance characteristic of large-scale motions (right frame of figure \ref{fig:pmspectra_yp15_y05_Retau180}), the peak is located at $(\lambda_x,~\lambda_z) \approx (1.40,~2.09)$.
The spectra without premultiplication are shown in figure \ref{fig:spectra_y0_y05_Retau180} shows 
At both the wall and $y \approx 0.5$, the spectra peak at $(\alpha,~\beta) = (0.5,~0)$, showing spanwise elongated pressure structures.
A peak of the pressure spectrum for $\beta \to 0$ was also documented in \citet{yang2022wavenumber}.

\begin{figure}[!h]
	\centerline{\includegraphics[width=\textwidth]{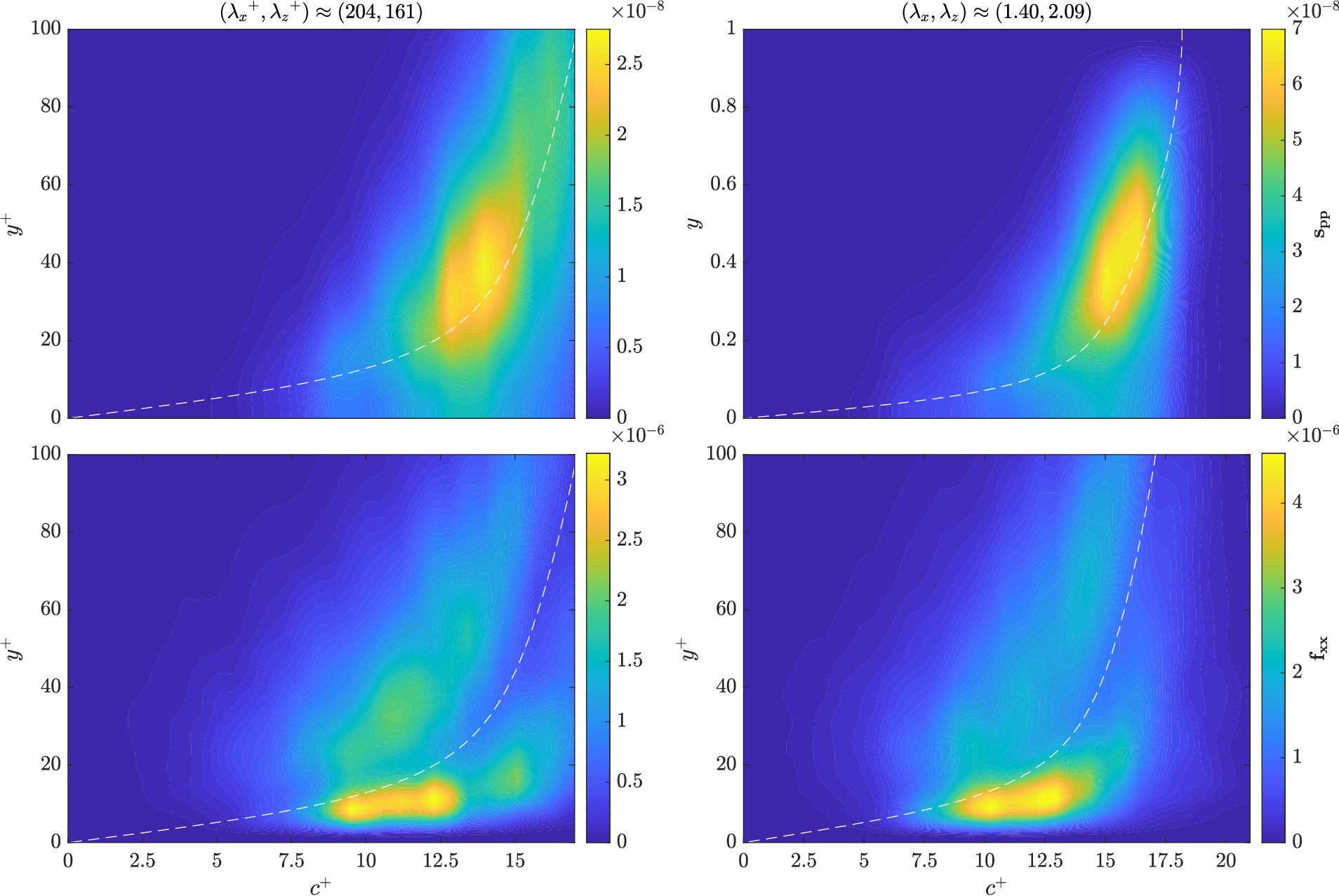}}
	\caption{PSDs for $Re_{\tau} \approx 180$ simulation. Left frames: near-wall structures. Right frames: large-scale structures. Top frames: pressure component ($\boldsymbol{s}_{pp}$). Bottom frames: streamwise forcing component ($\boldsymbol{f}_{xx}$). Dashed lines denote the critical layer.}
	\label{fig:psd_y0_y05_Retau180}
\end{figure}

The pressure and forcing components PSDs are shown in figure \ref{fig:psd_y0_y05_Retau180} for the near-wall and large-scale structures.
For the near-wall structures, the pressure component PSD peaks at $c^+ \approx 14$ ($\omega^+ \approx 0.43$), with spatial support around $y^+ \approx 38$, whereas for the large-structures, such peak is located at $c^+ \approx 16$ ($\omega \approx 4.52$), with support $y \approx 0.42$.
These peaks arise along the critical layer, shown in the plots with dashed lines.
The $\omega^+$ peak value is roughly in agreement with previous results \citep{hu2006wall, anantharamu2020analysis}.
Regarding the streamwise forcing component PSD, the near-wall and large-scale structures peak at $c^+ \approx 12.5$, with support close to the wall, at $y^+ \approx 10$.

\begin{figure}[!h]
	\centerline{\includegraphics[width=\textwidth]{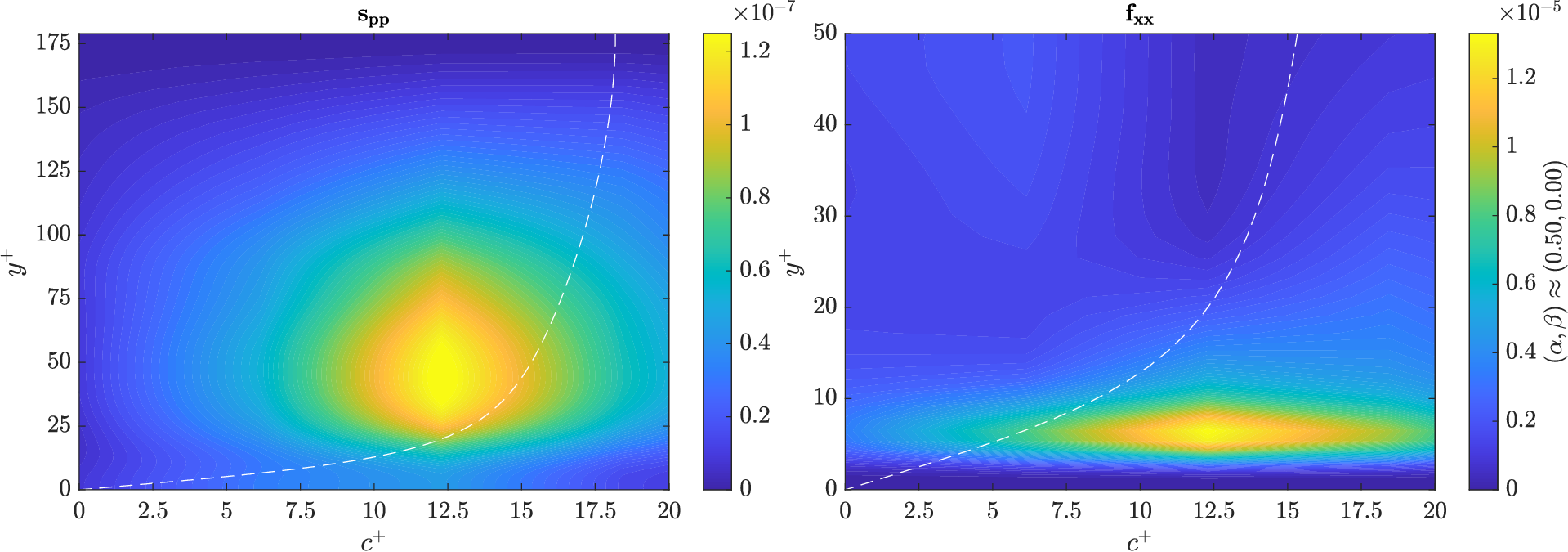}}
	\caption{PSDs for $Re_{\tau} \approx 180$ simulation and spanwise-coherent structure, i.e. $(\alpha,~\beta) = (0.5,~0)$. Right frame: pressure component ($\boldsymbol{s}_{pp}$). Left frame: streamwise forcing component ($\boldsymbol{f}_{xx}$). Dashed lines denote the critical layer.}
	\label{fig:psd_PSP_Retau180}
\end{figure}

Figure \ref{fig:psd_PSP_Retau180} shows the pressure (right frame) and streamwise forcing (left frame) components PSDs for the peak at $(\alpha,~\beta) = (0.5,~0)$, which is present for the power spectra without premultiplication (figure \ref{fig:spectra_y0_y05_Retau180}).
Both PSDs peak at $c^+ \approx 12.5$, although for the pressure component the peak has support at $y^+ \approx 44$, whereas for the streamwise forcing component the support is much closer to the wall, at $y^+ \approx 6.5$.

\begin{figure}[!h]
	\centerline{\includegraphics[width=\textwidth]{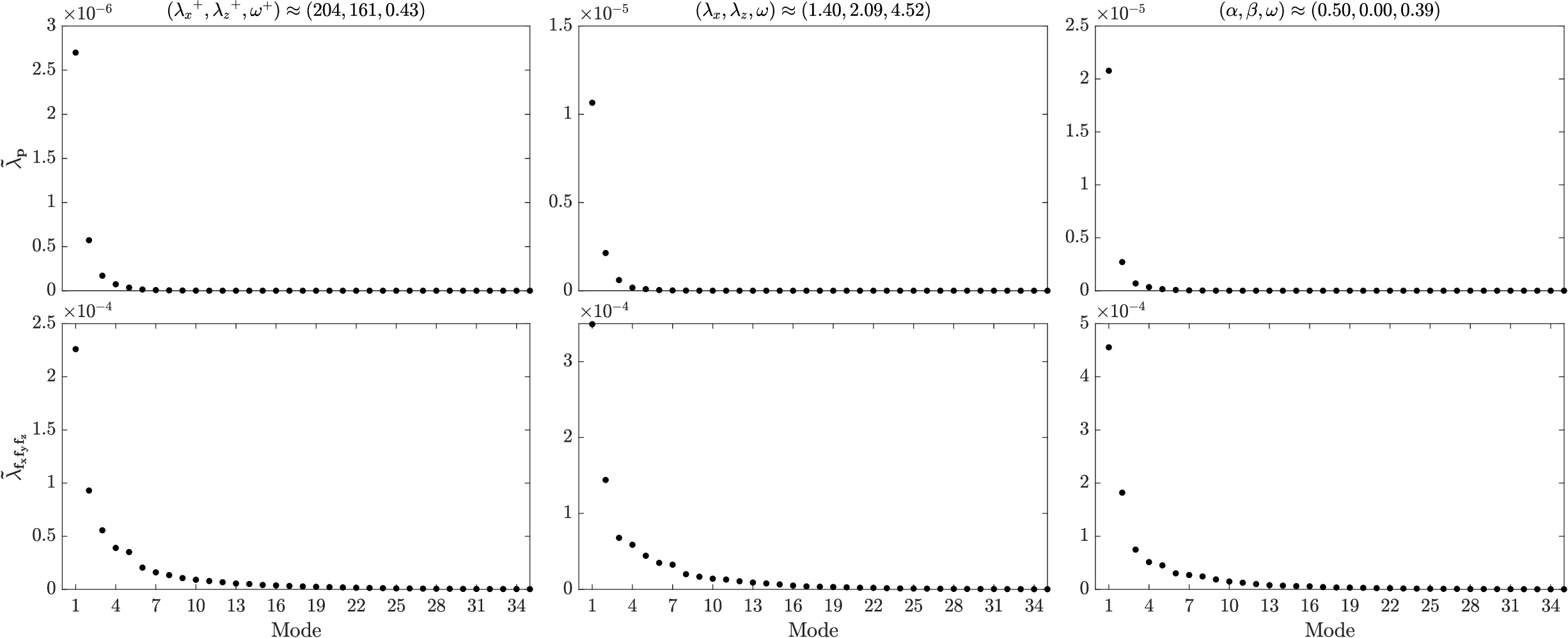}}
	\caption{SPOD eigenvalues employing the pressure (top frames) and forcing (bottom frames) modes for the $Re_{\tau} \approx 180$ simulation. See the comments in the caption of figure \ref{fig:spod_p_fxfyfz_eigenvalues_Retau550}.}
	\label{fig:spod_p_fxfyfz_eigenvalues_Retau180}
\end{figure}

\begin{figure}[!h]
	\centerline{\includegraphics[width=\textwidth]{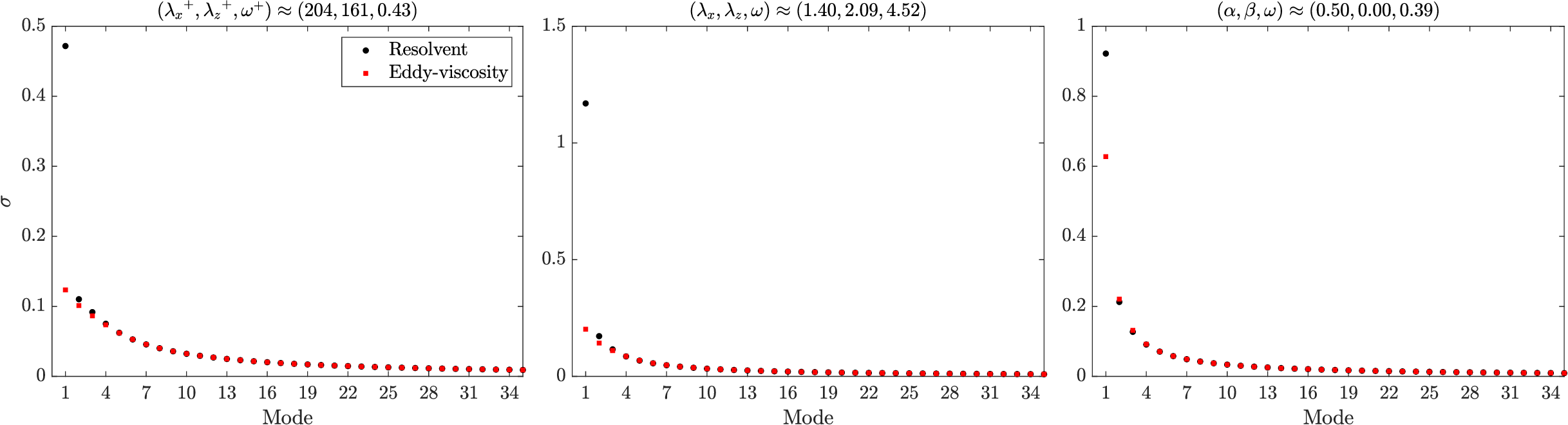}}
	\caption{Resolvent gains employing the pressure weighing for the $Re_{\tau} \approx 180$ simulation. See the comments in the caption of figure \ref{fig:resolvent_p_gains_Retau550}.}
	\label{fig:resolvent_p_gains_Retau180}
\end{figure}

The SPOD eigenvalues and the resolvent eigenvalues are shown in figures \ref{fig:spod_p_fxfyfz_eigenvalues_Retau180} and \ref{fig:resolvent_p_gains_Retau180}.
Similar results as those obtained for the $Re_{\tau} \approx 550$ simulation are observed for the $Re_{\tau} \approx 180$ simulation, i.e. the SPOD eigenvalue and the resolvent gains without the eddy viscosity present low-rank behavior, whereas the eddy viscosity resolvent gains do not.

\begin{figure}[!h]
	\centerline{\includegraphics[width=\textwidth]{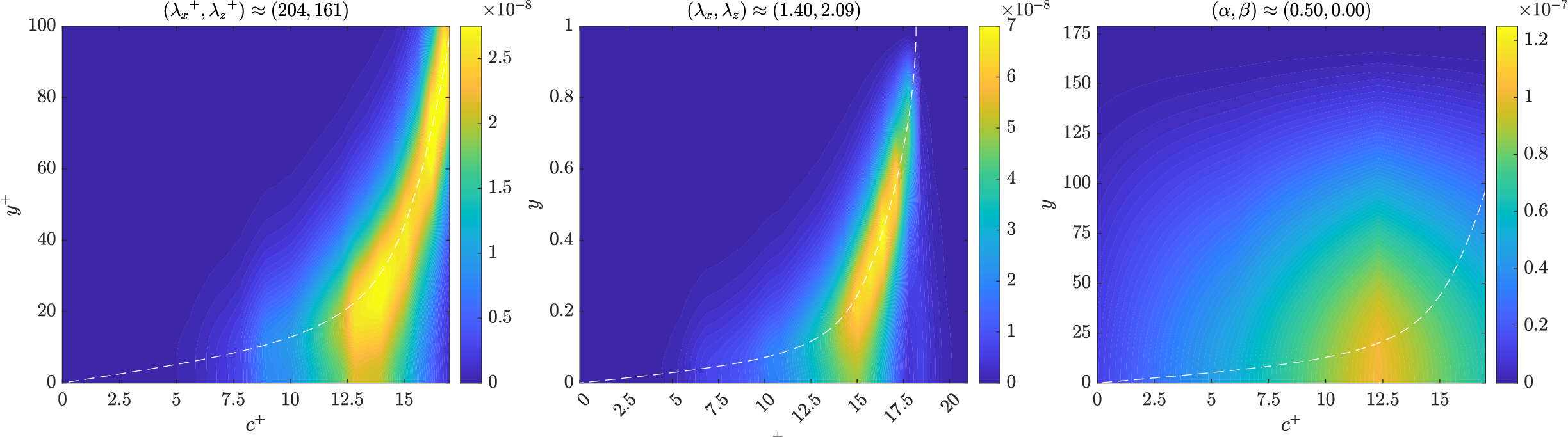}}
	\caption{Pressure component PSD low-rank resolvent reconstruction (1st mode) for the $Re_{\tau} \approx 180$ simulation. See the comments in the caption of figure \ref{fig:resolvent_PSD_reconstruction_Retau550}.}
	\label{fig:resolvent_PSD_reconstruction_Retau180}
\end{figure}

\begin{figure}[!h]
	\centerline{\includegraphics[width=\textwidth]{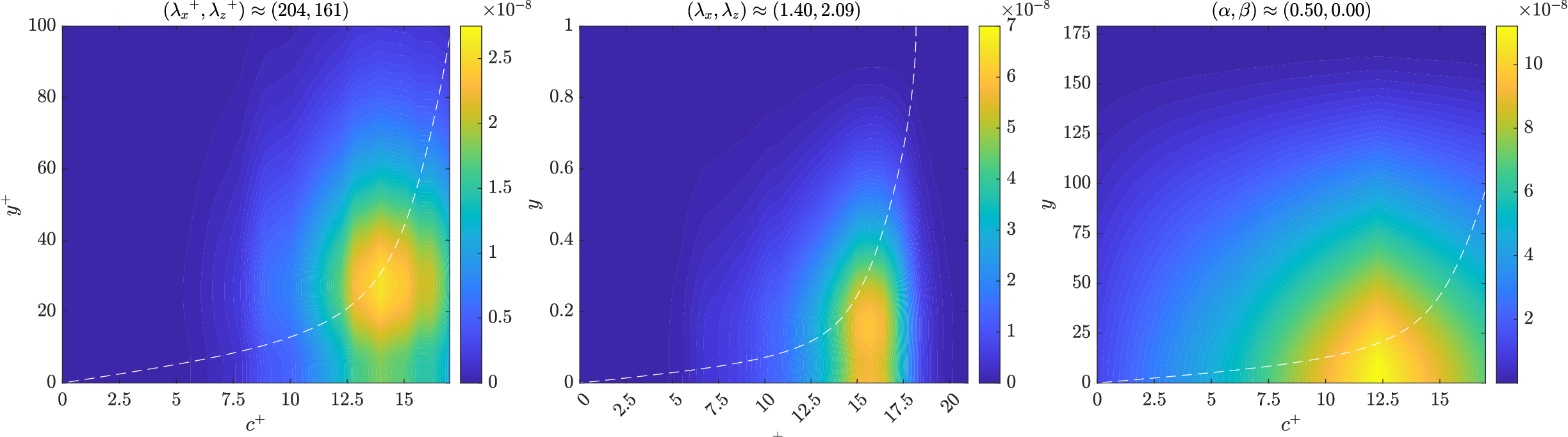}}
	\caption{Pressure component PSD low-rank eddy viscosity resolvent reconstruction (1st mode) for the $Re_{\tau} \approx 180$ simulation. See the comments in the caption of figure \ref{fig:resolvent_PSD_reconstruction_Retau550}.}
	\label{fig:resolventEV_PSD_reconstruction_Retau180}
\end{figure}

Low-rank representations of the pressure PSD using the first resolvent mode, without and with eddy viscosity, are depicted in figures \ref{fig:resolvent_PSD_reconstruction_Retau180} and \ref{fig:resolventEV_PSD_reconstruction_Retau180}.
As for the $Re_{\tau} \approx 550$ case, the low-rank model was able to capture the main features of the DNS PSD, especially for the resolvent without eddy viscosity.

\begin{figure}[!h]
	\centerline{\includegraphics[width=\textwidth]{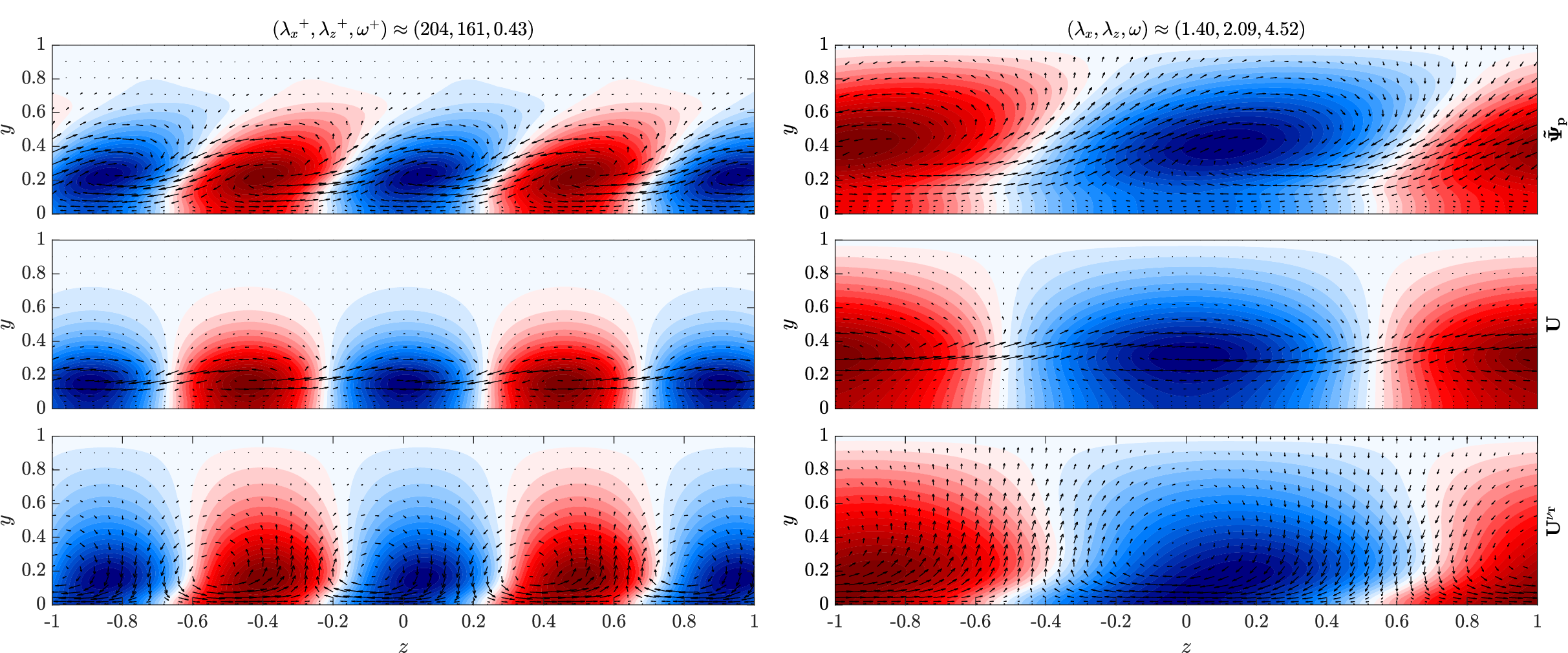}}
	\caption{Pressure near-wall (left frames) and large-scale (right frames) structures modes for the $Re_{\tau} \approx 180$ simulation. See the comments in the caption of figure \ref{fig:p_spod_resolvent_modes_near-wall_large-scale_Retau550}.}
	\label{fig:p_spod_resolvent_modes_near-wall_large-scale_Retau180}
\end{figure}

\begin{figure}[!h]
	\centerline{\includegraphics[width=\textwidth]{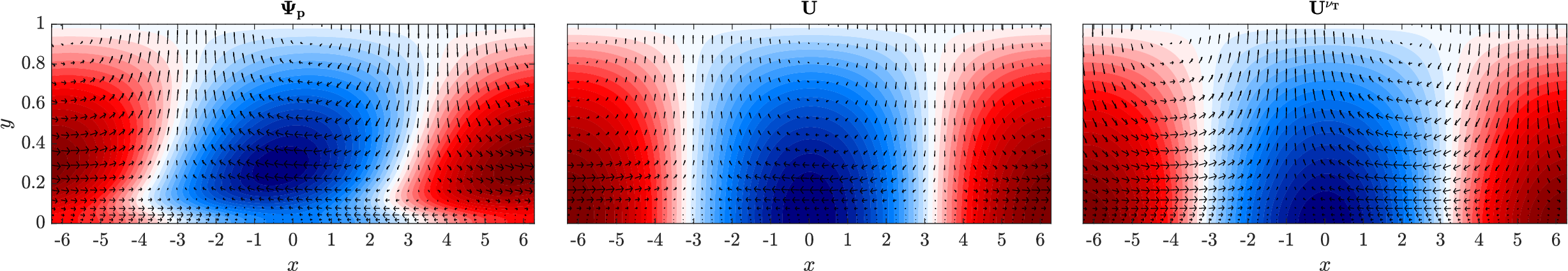}}
	\caption{Pressure spanwise-coherent structures for the $Re_{\tau} \approx 180$ simulation. See the comments in the caption of figure \ref{fig:p_spod_resolvent_modes_spanwise-coherent_Retau550}.}
	\label{fig:p_spod_resolvent_modes_spanwise-coherent_Retau180}
\end{figure}

Near-wall, large-scale, and spanwise-coherent pressure modes obtained through SPOD and resolvent analysis are shown in figures \ref{fig:p_spod_resolvent_modes_near-wall_large-scale_Retau180} and \ref{fig:p_spod_resolvent_modes_spanwise-coherent_Retau180}.
The results are similar to those obtained for the $Re_{\tau} \approx 550$ simulation (figures \ref{fig:p_spod_resolvent_modes_near-wall_large-scale_Retau550} and \ref{fig:p_spod_resolvent_modes_spanwise-coherent_Retau550}).
The streamwise component of the $\boldsymbol{\tilde{\Psi}}_{p}$ (SPOD) modes are inclined to the right for $y \geq 0.1$, whereas for the $\boldsymbol{U}$ (resolvent) modes the regions of positive and negative pressure are normal to the wall.
On the other hand, the streamwise component of the $\boldsymbol{U}^{{\nu_T}}$ (eddy viscosity resolvent) modes are inclined to the left close to the wall, up to $y \leq 0.25$, whereas for $y > 0.25$, they are slightly inclined to the right.
Such feature is evident for the near-wall and large-scale structures and at some extent for the spanwise-coherent structure.

\FloatBarrier


\bibliographystyle{unsrtnatabbrv}
\bibliography{references}

\end{document}